\documentclass[%
 reprint,
superscriptaddress,
 amsmath,amssymb,
 aps,
 pre,
longbibliography,
]{revtex4-1}
\usepackage[utf8]{inputenc}
\usepackage{upgreek}
\usepackage{amssymb}
\usepackage{amsmath}
\usepackage{verbatim}
\usepackage{graphicx}
\usepackage{textcomp}
\usepackage{booktabs}
\usepackage{bbold}
\usepackage{xcolor}
\usepackage{graphicx}% Include figure files
\usepackage{dcolumn}% Align table columns on decimal point
\usepackage{bm}% bold math
\usepackage[colorlinks,
            linkcolor=red,
            anchorcolor=blue,
            citecolor=blue
            ]{hyperref}

\begin{document}
\title{Quantum walks: the first detected transition time}

\author{Q. Liu}
\affiliation{Department of Physics, Institute of Nanotechnology and Advanced Materials, Bar-Ilan University, Ramat-Gan 52900, Israel}
\affiliation{School of Physics, Taishan College, Shandong University, Jinan 250100, China}

\author{R. Yin} 
\affiliation{Department of Physics, Institute of Nanotechnology and Advanced Materials, Bar-Ilan University, Ramat-Gan 52900, Israel}

\author{K. Ziegler}
\affiliation{Institut f\"ur Physik, Universit\"at Augsburg, $D-86135$ Augsburg, Germany}

\author{E. Barkai}
\affiliation{Department of Physics, Institute of Nanotechnology and Advanced Materials, Bar-Ilan University, Ramat-Gan 52900, Israel}

\date{\today}

\begin{abstract}
We consider the quantum first detection problem for a particle evolving on a graph under repeated projective measurements with 
fixed rate $1/\tau$. A general formula for the mean first detected transition time 
is obtained for a quantum walk in a finite-dimensional Hilbert space where the initial state 
$|\psi_{\rm in}\rangle$ of the walker is orthogonal to the detected state $|\psi_{\rm d}\rangle$. We focus on diverging mean transition times, where the total detection 
probability exhibits a  discontinuous drop of its value, by mapping the problem onto a theory of fields of classical charges located on the unit disk.  
Close to the critical parameter of the model, which exhibits a blow-up of the mean transition time, 
we get simple expressions for the mean transition time. Using previous results on the fluctuations of the return time,
corresponding to $|\psi_{\rm in}\rangle = |\psi_{\rm d}\rangle$, we find close to these critical parameters that the mean transition time is proportional to the fluctuations of the return time, an expression reminiscent  of the Einstein relation.
\end{abstract}

\maketitle

\section{\label{sec:level1}Introduction}
\begin{figure}
    \centering
    \includegraphics[width=0.99\columnwidth]{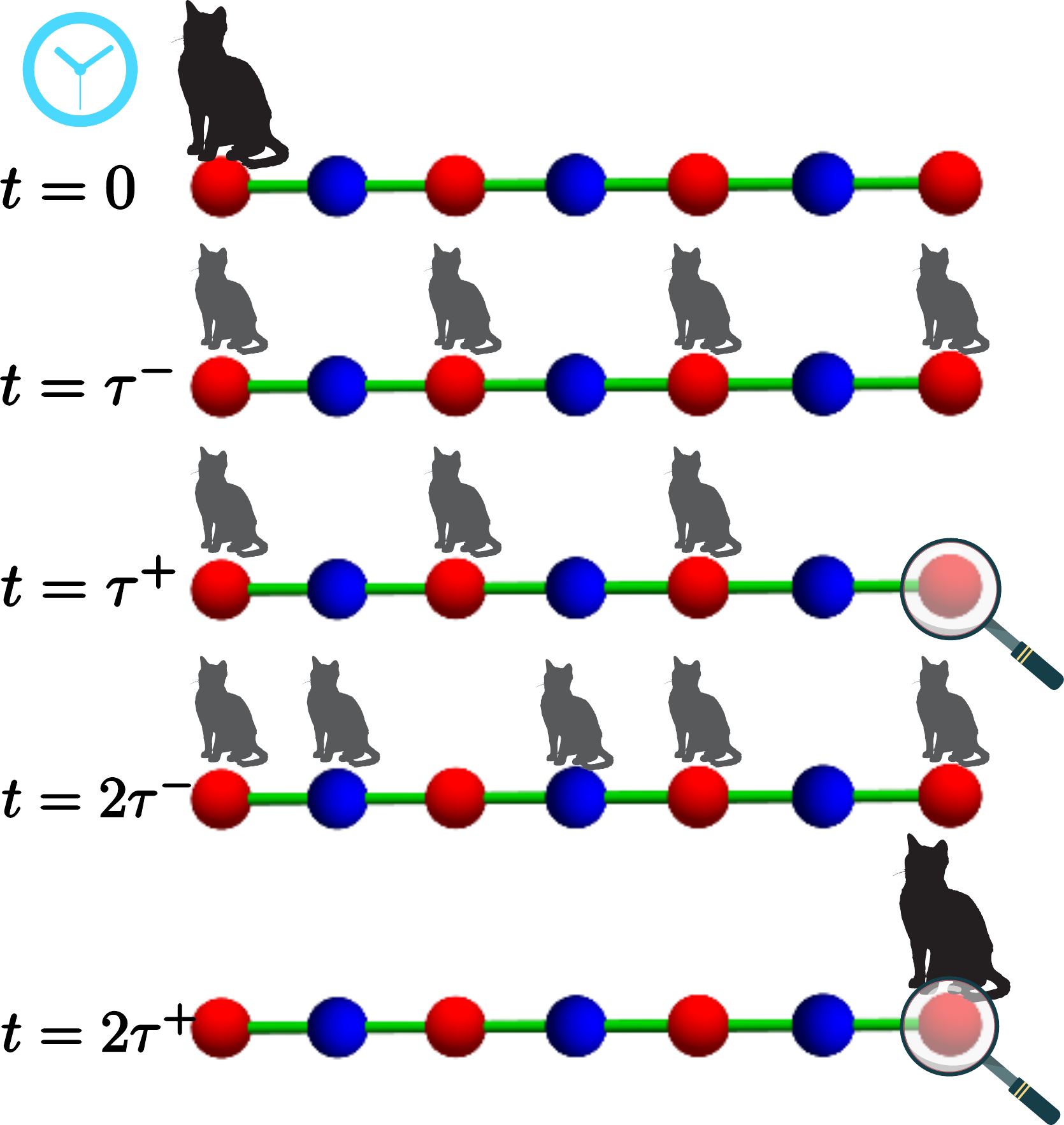}
    \caption{Schematic plot of the first detected transition problem in quantum walks. The quantum particle is prepared at the initial state (black cat) at $t=0$ and evolves unitarily (gray cats) in the detection free interval $\tau$. And measurements (the magnifying glass) are performed every $\tau$ units of time. Here $-$/$+$ means before/after measurement. In the failed attempts $t=\tau$, the detector collapses the wave function at the detected state. We repeat this process until the quantum particle is successfully detected (for example, here $t=2\tau$). The question is how long it takes to find the quantum particle.}
    \label{fig:quantumwalks}
\end{figure}
A closed quantum system is prepared in some initial state and evolves unitarily over time. Our aim is to monitor the evolution of this 
system by repeated projective measurements until a certain state is detected for the first time. A corresponding simple classical \cite{Redner2001,ralf2014first} example
would be to take a picture of a rare animal in the wilderness. For this purpose a remote camera takes pictures at a fixed rate, and the camera's 
software checks immediately whether the rare animal is on the last picture or not. Once the animal is caught on the last snapshot the process stops. 
It is obvious that we may miss the first appearance of the animal in the process. But when we continue long enough we might be lucky. 
The theoretical question is, what would be ``long enough" to detect the animal at a given measurement rate?

Quantum walks are well investigated both theoretically and experimentally \cite{Aharonov1993,Farhi1998,Childs2002,Karski174,PhysRevLett.104.100503,Preiss1229}.  
Also the quantum first detection problem, for a quantum walk on a graph has been considered in detail \cite{BACH2004562,Krovi2006,PhysRevA.74.042334,Varbanov2008,Gruenbaum2013,Grunbaum2014,Krapivsky2014,Dhar2015,Dhar_2015,Friedman_2016,Friedman2017,Thiel2018}
, as part of a wider investigation of unitary evolution pierced by measurements \cite{PhysRevB.98.104309,PhysRevE.98.022129,belan2019optimal,PhysRevA.99.062105,benzion2019disentangling,zabalo2019critical,PhysRevX.9.031009,roy2019measurementinduced}.
The rate $1/\tau$ at which we detect the particle on a given site becomes a crucial parameter, for example, if we sample too fast the ``animal'' 
cannot be detected at all due to the Zeno effect. This implies that there exist special sampling times that are optimal, in the sense that the 
detection time attains a minimum. Indeed it was shown by Krovi and Brun \cite{Krovi2006,PhysRevA.74.042334,Varbanov2008} that on certain graphs,
due to constructive interference, the quantum search problem is highly efficient. At the same time, these authors noted that in other cases, 
 destructive interference may render the quantum search inefficient in the sense that the hitting time even for a small system can be infinity 
(unlike classical random walks on a finite graph). In this paper we use a recently proposed quantum renewal equation  \cite{Friedman2017} to find the average time 
of a quantum walker starting on $|\psi_{\rm in}\rangle$ to be detected on $|\psi_{\rm d}\rangle$.

We employ stroboscopic sampling, which allows for considerable theoretical advance, with generating function technique.  
It is hoped  that in the long run, this type of investigation will lead to advances in quantum search algorithms \cite{PhysRevLett.79.325,PhysRevA.70.022314,BACH2004562,Kempe,PhysRevLett.116.100501}.  More importantly, 
in this work we map the problem of calculating the averaged transition time to a classical charge theory. We show how the mean quantum transition time is related to the stationary points of a set of classical charges positioned on the unit circle in the complex
plane with locations $\exp(i E_j \tau)$. This charge picture was previously promoted in the context of the return problem \cite{Gruenbaum2013}
($|\psi_{\rm in}\rangle = |\psi_{\rm d}\rangle$), while here we use this method to solve the transition problem. 
These two problems exhibit vastly different behavior. For the return problem the mean return time is quantized, since it is given as a topological
invariant which is the winding number of the wave function \cite{Gruenbaum2013,Yin2019}. In our problem this is equal to the dimensionality of the 
underlying Hilbert space with non-degenerate eigenvalues of the back-folded spectrum. 
Thus, the average return time is independent of the sampling rate. 
In contrast, the transition time is very sensitive, for instance, to the sampling rate $1/\tau$, and its behaviors are 
highly non-trivial \cite{Friedman2017}. 

The rest of this paper is organized as follows: In Sec. \ref{model and formalism} we define our model and the degeneracies
caused by the sampling time $\tau$. Then we derive our first main result the mean first detected transition (FDT) time in Sec. 
\ref{First detected transition time}. We find the general relation of the transition time and return time fluctuations in 
Sec. \ref{The detected transition time and return fluctuations}. In Secs. \ref{Weak charge},\ref{Two merging charges},\ref{time and fluctuation relation},
\ref{Big charge theory} we study some characteristic diverging transition times, where special relations for the transition 
time and the return fluctuations are found. This includes some examples to confirm our theory. We close the paper with 
discussions and a summary in Sec. \ref{discussion}. Detailed calculations are presented in the appendices.

\section{Model and Formalism\label{model and formalism}}

\subsection{Stroboscopic Protocol}
We consider a quantum particle prepared in the state $|\psi_{\rm in}\rangle$, for instance on a node of the 
lattice or other graphs. The evolution of this quantum particle is described by the time-independent Hamiltonian $H$ 
according to the Schr\"{o}dinger equation. As an example consider a one-dimensional tight-binding model in discrete position space 
with nearest neighbor hopping:
\begin{equation}
    H=-\gamma \sum_{x=-N}^{N}(|x\rangle\langle x+1|+|x+1\rangle\langle x|).
\end{equation}
However, our general formalism does not rely on a specific Hamiltonian, as long as we are restricted 
to a finite-dimensional Hilbert space. 

In a measurement the detector collapses the wave function at the detected state 
$|\psi_{\rm d}\rangle$ by the projection operator $D=|\psi_{\rm d}\rangle\langle\psi_{\rm d}|$. For simplicity one may 
assume that $|\psi_{\rm d}\rangle$ is yet another localized node state of the graph, however our theory is developed in generality.  
We perform the measurements with a discrete time sequence $\tau,2\tau,\cdots,n\tau$ until it is successfully detected for the first time. Then the 
result of the measurements is a string: ``no, no, $\cdots$,no, yes''. In the failed measurements the wave function collapses to zero at the 
detected state, and we renormalize the wave function after each failed attempt. The event of detecting the state 
$| \psi_{\rm d}\rangle$ for the first time after $n$ attempts implies that $n-1$ previous attempts failed and this 
certainly must be taken into consideration. Namely the failed measurements back fire and influence the dynamics, by 
erasing the wave function at the detected state.  Finally, the quantum state is detected and the experiment is 
concluded (see Fig. \ref{fig:quantumwalks}). Hence the first detection time is $t= n \tau$.

The key ingredients of this process are the initial state $|\psi_{\rm in}\rangle$ and the detected state $|\psi_{\rm d}\rangle$, 
which characterize this repeated measurements problem. If the initial state is the same as the detected state, namely $\langle\psi_{\rm in}|\psi_{\rm d}\rangle=1$ 
we call this case the first detected return (FDR) problem, which has been well studied by a series of works 
\cite{Gruenbaum2013,Stefanak2008,Xue2015,Dhar2015,Yin2019}. In the following we investigate the FDT problem,
where $\langle\psi_{\rm in}|\psi_{\rm d}\rangle=0$. This transition problem describes the transfer of the quantum state from $|\psi_{\rm in}\rangle$ 
to $|\psi_{\rm d}\rangle$ in Hilbert space. The time this process takes is of elementary importance. Since the results in each experiment are random,
we focus on the expected FDT time $\langle n \rangle\tau$, which gives the average quantum transition time in the presence of the stroboscopic measurements.

During each time interval $\tau$ the evolution of the wave function is unitary $|\psi(n\tau^-)\rangle=U(\tau)|\psi[(n-1)\tau^+]\rangle$,
where $U(\tau)=\exp{(-i H \tau)}$ (we set $\hbar=1$ in this paper) and $-$/$+$ means before/after measurement.  Let $\phi_n$ be the FDT amplitude, 
the probability of the FDT in the $n$-th measurement is $F_n=|\phi_n|^2$. If the particle is detected with probability one (see further details 
\cite{Thiel2019}), which means $\sum_{n=1}^{\infty}|\phi_n|^2=1$, the mean FDT time is $\langle t \rangle=\tau\sum_{n=1}^{\infty}n|\phi_n|^2$. 
As we will soon recap, $\phi_n$ can be evaluated from a unitary evolution interrupted by projective measurements. However, there exist a deep relation 
between $\phi_n$ and the unitary evolution without interrupting measurement.

\subsection{Brief summary of the main results}

Before we start with the general discussion of the evolution of a closed quantum system under repeated measurements,
we would like to summarize the main results:
Repeated measurements interrupt the unitary evolution by a projection after a time step $\tau$. This has a strong effect on the
dynamical properties, which can be observed in the transition amplitude $\phi_{n}$ of Eq. (\ref{phi origen}). 
The unitary evolution $\exp{(-i H \tau)}$ is controlled by the energy spectrum. The overlaps $\{p_k\}$ and $\{q_k\}$ in 
Eqs. (\ref{qk},\ref{pk}) are crucial in that they connect the eigenstates of $H$ and the initial and measured states.
The non-unitary evolution is characterized by the zeros of the polynomial Eq. (\ref{D}) and these overlap functions. Those zeros are
formally related to a classical electrostatic problem \cite{Gruenbaum2013}; namely they are the stationary points of a test charge in a system with charges on the unit circle, which is defined in Eq. (\ref{Force field}). After solving this electrostatic problem, the zeros are used to calculate, for instance, the first detection amplitude with Eq. (\ref{phi n}), the divergent behavior of the mean FDT
time near degenerate points in Eq. (\ref{mean n excat}), and a generalized Einstein relation between the mean FDT time and the 
FDR variance in Eqs. (\ref{n and var},\ref{einstein1}). This leads us to the conclusion that the mean FDT time, i.e. the mean time to reach a certain quantum state, is very sensitive to the time step $\tau$ of the measurements. In particular, degeneracies of the back-folded spectrum
in Eq. (\ref{back_folded_spectrum}) can lead to extremely long times for the detection of certain quantum states.
Based on this general approach, we have calculated the mean FDT for a two-level system in Eq. (\ref{n two-level}),  
for a Y-shaped molecule in Eq. (\ref{n Y shape}), and for a Benzene-type ring in Sec. \ref{sect:ring}.

\subsection{Generating function}
The FDT amplitude $\phi_{t,n}$ for the evolution from $| \psi_{\rm in}\rangle$ to $| \psi_{\rm d}\rangle$ and the 
FDR amplitude $\phi_{r,n}$ for the evolution from $| \psi_{\rm d}\rangle$ to $| \psi_{\rm d}\rangle$ read \cite{Gruenbaum2013,Dhar2015,Friedman_2016,Friedman2017}
\begin{equation}
    \phi_{t,n}=\langle \psi_{\rm d}| (e^{-i \tau H }P)^{n-1}e^{-i \tau H}| \psi_{\rm in}\rangle,
    \label{phi origen}
\end{equation}
\begin{equation}
    \phi_{r,n}=\langle \psi_{\rm d}| (e^{-i \tau H }P)^{n-1}e^{-i \tau H}| \psi_{\rm d}\rangle,
    \label{phi origen r}
\end{equation}
with $P=1-D=1-|\psi_{\rm d}\rangle\langle\psi_{\rm d}|$. As the equations show, the unitary evolution in the detection free interval $\tau$ 
is interrupted by the operation $P$. The combined unitary evolution and the projection goes with the power $n-1$,
corresponding to the $n-1$ prior failed measurements. 
Moreover, we define the unitary transition amplitude $v_n$ and the unitary return amplitude $u_n$ as
\begin{equation}
    v_n=\langle \psi_{\rm d} | e^{-i n H \tau} | \psi_{\rm in}\rangle ,
    \label{vn}
\end{equation}
\begin{equation}
    u_n=\langle \psi_{\rm d}| e^{-i n H \tau}| \psi_{\rm d}\rangle .
    \label{un}
\end{equation}
These amplitudes describe transitions from the initial state to the detected state and from the detected state back to
itself, free of any measurement. Using the $v_n$ and $u_n$, we expand  Eq. (\ref{phi origen}) and (\ref{phi origen r}) 
in $P$ which leads to an iteration equation known as the quantum renewal equation \cite{Friedman_2016,Friedman2017}:
\begin{eqnarray}
      \phi_{t,n} &=& v_n-\sum_{j=1}^{n-1}\phi_{t,j} u_{n-j},\label{quantum renewal}\\
      \phi_{r,n} &=& u_n-\sum_{j=1}^{n-1}\phi_{r,j} u_{n-j}.\label{quantum renewal r}
\end{eqnarray}
Note that the first terms $v_n$, $u_n$ on the right-hand side describe the unitary evolution between the initial state and the detected state
and between the detected state to itself. The second terms describe all the former wave function returns to the detected state. These 
recursive equations, together with the exact function Eq. (\ref{mean n excat}) for mean transition times, are used in the example section
to find exact solutions of the problem. In order to solve the recursive equations a direct method is to transform the quantum renewal equation into the frequency (or $\omega$) space. 
Since the renewal equations consist of $\{ v_j\}$ and $\{ u_j\}$, we need to transform these quantities into $\omega$ space first. 
Using Eqs. (\ref{vn},\ref{un}) we have
\begin{equation}
    \hat{v}( \omega):=\sum_{n=1}^{\infty}e^{i n \omega}v_n=\langle \psi_{\rm d}|(e^{i \tau H-i \omega}-1)^{-1}|\psi_{\rm in}\rangle,
\end{equation}

\begin{equation}
    \hat{u}( \omega):=\sum_{n=1}^{\infty}e^{i n \omega}u_n=\langle \psi_{\rm d}|(e^{i \tau H-i \omega}-1)^{-1}|\psi_{\rm d}\rangle.
\end{equation}
The analogous calculation for the amplitudes $\phi_{t,n}$, $\phi_{r,n}$ leads to
\begin{equation}
    \hat{\phi}_t(\omega)\equiv\sum_{n=1}^{\infty}e^{i \omega n}\phi_{t,n}=\langle \psi_{\rm d}| A_{\omega}| \psi_{\rm in}\rangle,
\end{equation}
\begin{equation}
    \hat{\phi}_r(\omega)\equiv\sum_{n=1}^{\infty}e^{i \omega n}\phi_{r,n}=\langle \psi_{\rm d}| A_{\omega}| \psi_{\rm d}\rangle.
\end{equation}
where $A_{\omega}=(e^{i \tau H-i \omega}-P)^{-1}$. The initial state $|\psi_{\rm in}\rangle$ distinguishes the return and transition problem. $A_{\omega}$ is related to the Green's function $(e^{i H\tau}/z-P)^{-1}$ of the non-unitary evolution \cite{Thiel2019a}. Its poles are the solutions of $\text{det}(\mathbb{1}/z-P U(\tau))=0$. We will see later that these poles are essential for the evaluation of the mean FDT time.  This property implies that the repeated measurement protocol can be possibly related to open quantum systems, in the sense that the measurements acting on the system is equivalent to the interaction between environment and the system \cite{PhysRevLett.123.140403,s2019complex}. Thus we believe that further research on this topic is worth while.

Using the identity $(1+B)^{-1}=1-B(1+B)^{-1}$, we obtain
\begin{equation}
    \langle \psi_{\rm d}| A_{\omega}| \psi_{\rm in}\rangle=\hat{v}( \omega)-\hat{u}( \omega)\langle \psi_{\rm d}| A_{\omega}| \psi_{\rm in}\rangle,
\end{equation}
\begin{equation}
    \langle \psi_{\rm d}| A_{\omega}| \psi_{\rm d}\rangle=\hat{u}( \omega)-\hat{u}( \omega)\langle \psi_{\rm d}| A_{\omega}| \psi_{\rm d}\rangle.
\end{equation}
Then the generating functions for the amplitude $\phi_t$ and $\phi_r$ read
\begin{equation}
    \hat{\phi}_t(\omega)=\frac{\hat{v}(w)}{1+\hat{u}(w)},\quad \hat{\phi}_r(\omega)=\frac{\hat{u}(w)}{1+\hat{u}(w)}.
\end{equation}
In the return problem, the initial state and detected state coincide, so the generating function only contains $\hat{u}(w)$. 
Whereas in the transition problem the symmetry is broken leading to the term $\hat{v}(w)$ in the numerator. 

A continuation of the phase factor $\exp{(i\omega)}$ from the unit disk to
the parameter $z$ in the complex plane is convenient for further calculations.
This leads to \cite{Friedman2017}
\begin{equation}
    \hat{\phi}_t(z)=\frac{\langle \psi_{\rm d}| \hat{U}(z)| \psi_{\rm in}\rangle}{1+\langle \psi_{\rm d}| \hat{U}(z)| \psi_{\rm d}\rangle},
    \label{generating}
\end{equation}
\begin{equation}
    \hat{\phi}_r(z)=\frac{\langle \psi_{\rm d}| \hat{U}(z)| \psi_{\rm d}\rangle}{1+\langle \psi_{\rm d}| \hat{U}(z)| \psi_{\rm d}\rangle},
    \label{generating r}
\end{equation}
where $\hat{U}(z)=\sum_{n=1}^{\infty}z^n U(n\tau)=z U(\tau) /(\mathbb{1}-z U(\tau))$ is the Z (or discrete Laplace) transform of $U(n\tau)$. 
The difference between Eq. (\ref{generating}) and Eq. (\ref{generating r}) is again only the numerator.

\subsection{Pseudo Degeneracy}
\label{sect:p_degeneracy}
The degeneracy of the energy levels plays a crucial role in the problem. For instance, a geometric symmetry of the graph can introduce such degeneracies. 
What is special here is that the measurement period $\tau$ leads to a new type of degeneracy of the distinct energy levels. This degeneracy is rooted 
in the stroboscopic sampling under investigation.

\begin{figure}
    \centering
    \includegraphics[width=0.8\columnwidth]{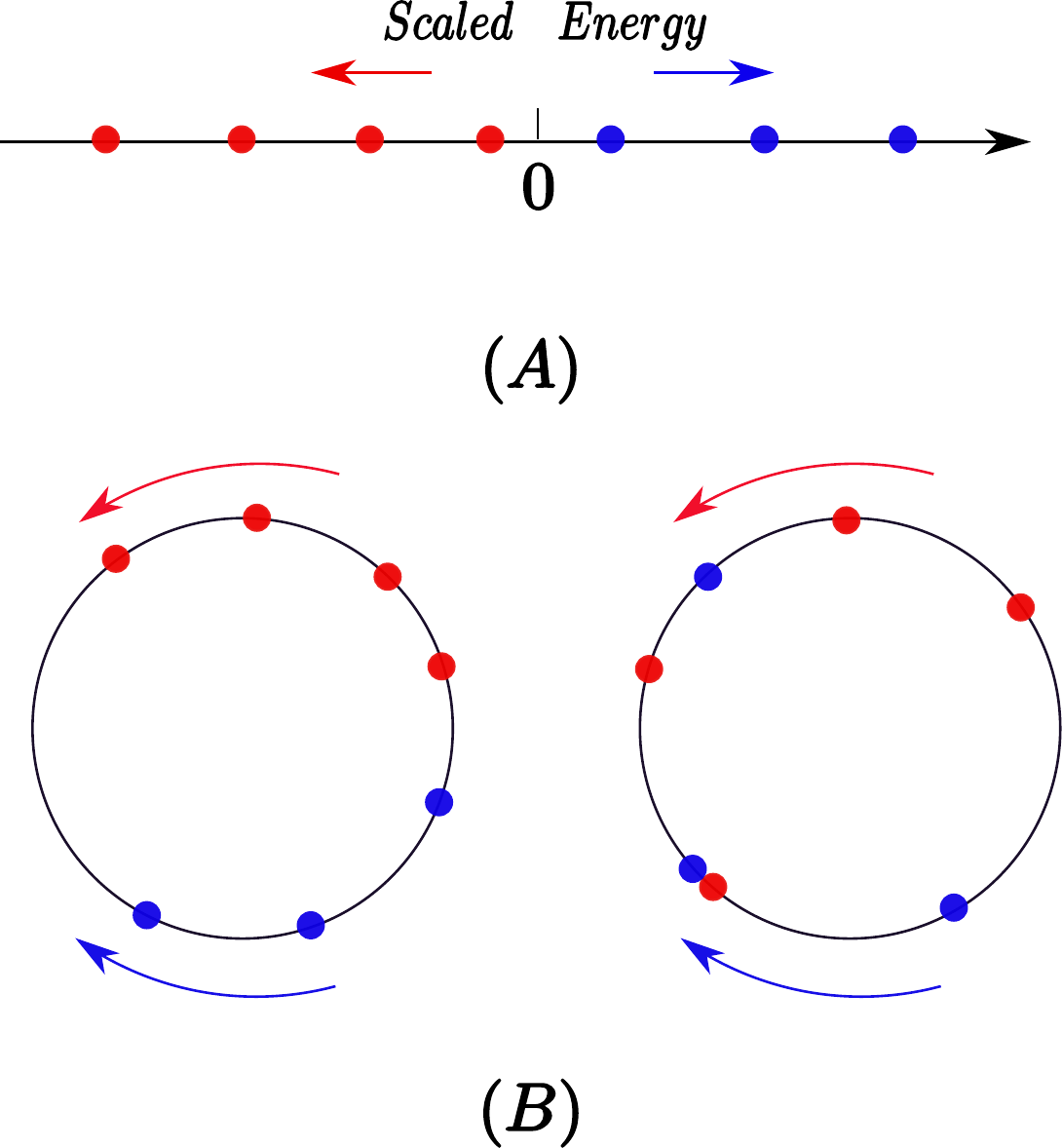}
    \caption{Schematic behaviors of (A) the scaled Hamiltonian spectrum $E_k\tau$ and (B) the phase $e^{-i E_k\tau}$ under a change of the sampling time $\tau$.
     The arrows indicate the movements of the scaled energy levels ($E_k\tau$) when increasing $\tau$. The positive (blue dots) and the negative (red dots) energy levels are
     well separated in (A). After mapping to the unit circle $E_k\tau\rightarrow e^{-i E_k \tau}$ they are not separated all the time, moving on the 
     unit circle making fusion of the phases possible. In particular, the case (right) can lead to degeneracies in the back-folded spectrum and to very large mean transition times.}
    \label{fig:energy mapping}
\end{figure}

For an arbitrary Hamiltonian $H$ which has $w$ non-degenerate energy levels, the eigenvalues $\{E_k\}_{k=0,...,w-1}$ of the Hamiltonian $H$ and the corresponding eigenstates  $\{| E_{k l}\rangle\}_{k=0,...,w-1}$ with $1 \leqslant l \leqslant g_k$, where $g_k$ is the degeneracy, can be used to express the matrix elements of Eq. (\ref{vn}) and Eq. (\ref{un})
in spectral representation as
\begin{eqnarray}
    v_n &=& \sum_{k=0}^{w-1}\bigg\{\sum_{l=1}^{g_k}\langle \psi_{\rm d}| E_{k l}\rangle\langle E_{k l}| \psi_{\rm in}\rangle\bigg\} e^{-i n E_k \tau},\\
    u_n &=& \sum_{k=0}^{w-1}\bigg\{\sum_{l=1}^{g_k}|\langle \psi_{\rm d}| E_{k l}\rangle|^2\bigg\} e^{-i n E_k \tau}.
\end{eqnarray}
These expressions are invariant under the change $ E_k\tau \rightarrow E_k\tau+2\pi j $ for integer $ j $. Thus, the eigenvalues 
$E_k$, $ E_{k^{\prime}} $ are effectively degenerate if $E_k\tau= E_{k^{\prime}}\tau+2\pi j $. Therefore, rather than the 
scaled eigenvalues $\{E_k \tau\}$ (which will be called simply eigenvalues subsequently), the back-folded eigenvalues $\{\Bar{E_k}\tau\}$
\begin{equation}
    \Bar{E_k}\tau=E_k\tau (\text{mod}\quad 2\pi) \quad -\pi \leqslant \Bar{E_k}\tau < \pi,
    \label{back_folded_spectrum}
\end{equation}
determine the dynamics at fixed $\tau$. This can also be understood as the mapping $E_k\tau\rightarrow e^{-i E_k\tau} $ from the real axis to the unit circle on the complex plane \cite{Gruenbaum2013} (see Fig. \ref{fig:energy mapping}). 
Here it is possible to change the value of $\tau$ until $\tau=\tau_c$ which leads to \cite{Gruenbaum2013,Friedman2017,Thiel2019}
\begin{equation}
   | E_k-E_i| \tau_c=2\pi j ,
   \label{excptional points}
\end{equation}
where $j$ is an integer. Thus, there are degeneracies of the back-folds eigenvalues for this critical $\tau_c$. 
Since the back-folded spectrum is relevant for the FDR/FDT and not the spectrum of $H$ itself, these degeneracies affect the discrete dynamics,
even if the eigenvalues $\{E_k\}$ of $H$ are non-degenerate. 

The quantum problem has a classical counterpart known as the first passage problem. The two problems  
exhibit vastly different behaviors, as might be expected. Let $P_{\rm det} =\sum_{n=1}^{\infty}|\phi_n|^2$ be the total detection probability. 
Unlike classical random walks on finite graphs, here one can find the total detection probability less than unity. The quantum particle will go to some 
``dark states'', where they will never be detected \cite{Thiel2019,Thiel2019a,Thiel2019b}. 

In Ref. \cite{Thiel2019,Thiel2019a} it was shown that $P_{\rm det}<1$ when the Hilbert space is split into two subspaces dark and bright.
The dark states can arise either
from degeneracies of the energy spectrum or from energy levels that have no overlap with the detected state.  The main focus of this paper is on 
cases where the total detection probability is unity (otherwise the search is clearly not efficient). Thus In our system we have $P_{\rm det}=1$,
except for special sampling times, given by Eq. (\ref{excptional points}). On these sampling times the detection probability is sub-optimal.
Close to these sampling times the mean time for detection will diverge, and one of our goals is to understand this behavior. 

\subsection{Zeros and Poles}

From $\hat{\phi}(z)=\sum_{n=1}^{\infty}z^n \phi_n$ we extract the amplitude $\phi_n$ by the inverse transformation \cite{Friedman2017}
\begin{equation}
    \phi_n=\frac{1}{2\pi i}\oint_{\gamma}\hat{\phi}(z)z^{-n-1}dz,
    \label{phi-n}
\end{equation}
where $\gamma$ is a counterclockwise closed contour around the circle of the complex plane with $|z|<1$,  where $\hat{\phi}(z)$ is analytic.
To perform the integration, we must analyze $\hat{\phi}(z)$. In Eqs. (\ref{generating},\ref{generating r}) the denominators only contain the state
$|\psi_{\rm d}\rangle$ and not the initial condition $|\psi_{\rm in}\rangle$, for both the FDR and FDT case.  
The poles outside the unit disc in turn will determine the relaxation pattern of $\phi_n$ (see below). To progress in our study of the transition problem we will 
use recent advances on the properties of the return problem \cite{Gruenbaum2013,Yin2019}. For this purpose we study the connection between the return and the transition problem more explicitly.
First, we define the overlap functions $p_k$ and $q_k$ of the initial/detected state as
\begin{eqnarray}
    q_k &=& \sum_{l=1}^{g_k}\langle \psi_{\rm d}| E_{kl}\rangle\langle E_{kl}| \psi_{\rm in}\rangle,\label{qk}\\
    p_k &=& \sum_{l=1}^{g_k}|\langle \psi_{\rm d}| E_{kl}\rangle|^2 ,
    \label{pk}
\end{eqnarray}
which correspond to the distinct energy level $E_k$ with degeneracy $g_k$. 
$q_k$ contains both detected and initial states while $p_k$ is only related to $|\psi_{\rm d}\rangle$.
These expressions indicate that $p_k$ is real and non-negative while $q_k$ is complex. The normalization of the energy eigenstates imply 
$\sum_{k=0}^{w-1}p_k=1$. On the other hand, $\sum_{k=0}^{w-1}q_k=0$, since the initial state and detected state are assumed to be orthogonal in the transition problem.

Next, we write the generating function in spectral representation as before, using eigenstates $|E_{kl}\rangle$ and the corresponding 
$g_k$-folded eigenvalues $E_k$. By  multiplying both  numerator and denominator $\prod_{k=0}^{w-1}(e^{i E_k \tau}-z)$, 
we express $\hat{\phi}_t(z)$ and $\hat{\phi}_r(z)$ as
\begin{equation}
    \hat{\phi}_t(z)=\frac{{\cal N}_t(z)}{{\cal D}(z)},\qquad \hat{\phi}_r(z)=\frac{{\cal N}_r(z)}{{\cal D}(z)}.
\end{equation}
Using $q_k$ and $p_k$ we can express ${\cal N}_t(z)$, ${\cal N}_r(z)$ and ${\cal D}(z)$ as
\begin{eqnarray}
     {\cal N}_t(z) &=& z\sum_{i=0}^{w-1} q_i \bigg[\prod_{k=0, k\neq i}^{w-1}(e^{i E_k \tau}-z)\bigg]\label{Nt},\\
     {\cal N}_r(z) &=& z\sum_{i=0}^{w-1} p_i \bigg[\prod_{k=0, k\neq i}^{w-1}(e^{i E_k \tau}-z)\bigg]\label{Nr},\\
     {\cal D}(z) &=& \sum_{i=0}^{w-1} p_i e^{i E_i \tau} \bigg[\prod_{k=0, k\neq i}^{w-1}(e^{i E_k \tau}-z)\bigg].
     \label{D}
\end{eqnarray}
The only difference between the ${\cal N}_r(z)$ and ${\cal N}_t(z)$ is that the $q_i$ in the former is replaced with $p_i$ in the latter. 
So $p_i$ and $q_i$ characterize the generating function of the return and the transition problem. ${\cal N}_r(z)$ and ${\cal D}(z)$ 
share the same multiplication term, each depending on the same group of real numbers $p_i$ and $\{E_i\}$. A straightforward calculation 
shows that the two polynomials are related \cite{Friedman2017}:
\begin{equation}
    {\cal D}(z)=(-1)^{w-1} e^{i\sum_j E_j\tau}z^w[{\cal N}_r(1/z^*)]^{\ast}.
    \label{relation D N}
\end{equation}{}

From Eqs. (\ref{generating},\ref{generating r}) the poles of the return and transition problem are identical. These poles, denoted by $Z_{i}$, 
are found from the solutions of ${\cal D}(Z)=0$. We also define the zeros of the generating function in the return problem, denoted by $z_{r,i}$. 
The latter are given by ${\cal N}_r(z)=0$. From Eq. (\ref{relation D N}), ${\cal D}(z)=(-1)^{w-1} e^{i\sum_j E_j\tau}z^w[{\cal N}_r(1/z^*)]^{\ast}=0$ 
yields $Z_{i}=1/z_{r,i}^*$. Hence transition poles $Z_{i}$ are given by
\begin{equation}
    Z_{i}=\frac{1}{z_{r,i}^{\ast}},\quad  z_{r,i}\neq 0.
    \label{relation of zero and pole}
\end{equation}{}
The key point is that the $\{Z_i\}$ describe both the transition problem investigated here and the return problem \cite{Gruenbaum2013}. 
Subsequently, we write $z_{r,i}$ as $z_{i}$ for simplicity. Eq. (\ref{relation of zero and pole}) gives us a way to find the poles $Z_i$
which are essential for the amplitude $\phi_n$, namely using the return zeros $z_i$, which have been studied already in the return problem 
\cite{Gruenbaum2013,Yin2019}.  

%The poles of the generating function are also related to the eigenvalues of the non-unitary evolution-eraser operator $s=(\mathbb{1}-D)U(\tau)$ \cite{Thiel2019a}. The characteristic polynomial of $s$ is $\text{det}(z\mathbb{1}-s)=\text{det}(z\mathbb{1}-U(\tau)+|d\rangle\langle d|U(\tau))$. Using the matrix determinant lemma together with Eq. (\ref{generating}), we obtain
%\begin{equation}
%    \text{det}(z\mathbb{1}-s)=\text{det}(z\mathbb{1}-U(\tau)){\cal D}(1/z).
%\end{equation}
%Hence the poles $\{Z_i\}$ are the reciprocal of the eigenvalues of the evolution-eraser operator $s$.  This property implies that the repeated measurement protocol can be possibly related to open quantum systems, in the sense that the measurements acting on the system is equivalent to the interaction between environment and the system \cite{PhysRevLett.123.140403,s2019complex,RevModPhys.70.101}. Thus we believe that further research on this topic is worth while.
%\begin{figure}
%    \centering
%    \includegraphics[width=0.99\columnwidth]{graph2/mappingthereturntotransition.pdf}
%    \caption{We show the zeros (green) and poles (red) of the return problem ($R$). They are related with $1/z*= %Z$.  
%    The poles of the transition problem ($T$) are identical to those of the return problem. If a zero of the return problem approaches   the unit disk so does the corresponding pole. In this case we will find large mean FDT times.}
%    \label{fig:zeros and poles}
%\end{figure}

\subsection{Charge Theory}
\label{sect:charge_th}

As already discussed before, the central goal is to determine the zeros $\{ z_i\}$.
A very helpful method in this regard was proposed by Gr\"{u}nbaum \textit{et al.} \cite{Gruenbaum2013}, 
who mapped the return problem to a classical charge theory. More importantly, the classical charge theory
provides an intuitive physical picture from which we can understand the behavior of the poles. 
Using Eq. (\ref{Nr}) for the zeros of ${\cal N}_r(z)$ with some rearrangement, we have $z\sum_{k=0}^{w-1}p_k/(e^{i E_k\tau}-z)=0$.
Neglecting the trivial zero at the origin we must solve
\begin{equation}
    {\cal F}(z)=\sum_{k=0}^{w-1}\frac{p_k}{e^{i E_k\tau}-z}=0.
    \label{Force field}
\end{equation}
${\cal F}(z)$ can be considered as a force field in the complex plane, 
stemming from charges $p_k$ whose locations are $e^{i E_k\tau}$ on the unit circle. 
Then the zeros $\{z_i\}$ of ${\cal N}_r(z)$ are the stationary points of this force field. 
Since there are $w$ charges which corresponds to the number of the discrete energy levels, 
we get $w-1$ stationary points in this force field from Eq. (\ref{Force field}). All the zeros are inside the unit disc, 
which is rather obvious since all the charges have the same sign ($p_k>0$). The physical significance of this is that the modes of the problem decay. 
More precisely, the zeros are within a convex hull, whose edge is given by the position of the charges, hence $|z_i|<1$. 
Then Eq. (\ref{relation of zero and pole}) implies $|Z_i|>1$, i.e. the poles lie outside the unit circle. 
%The correspondence of the return zeros and poles to transition zeros and poles is visualized in Fig. \ref{fig:zeros and poles}.

\section{FDT time\label{First detected transition time}}

In this section we focus on finding the general expression for the mean FDT time. 
We assume $\langle\psi_{\rm d}|\psi_{\rm in}\rangle=0$ which is the definition of ``transition". 
Since $\langle t \rangle=\tau\langle n \rangle=\tau\sum_{n=1}^{\infty}n|\phi_n|^2$, the first step 
is to find the amplitudes $\phi_n$, describing the detection probability for the $n$-th attempt. 
We start from the generating function of the FDT problem Eq. (\ref{generating}):
\begin{equation}
    \hat{\phi}_t(z)=\frac{z\sum_{i=0}^{w-1} q_i \bigg[\prod_{k=0, k\neq i}^{w-1}(e^{i E_k \tau}-z)\bigg]}
    {\sum_{i=0}^{w-1} p_i e^{i E_i \tau} \bigg[\prod_{k=0, k\neq i}^{w-1}(e^{i E_k \tau}-z)\bigg]},
\end{equation}
The numerator ${\cal N}_t(z)$ reads with the polynomial ${\cal G}(z)$   
\begin{equation} 
    {\cal N}_t(z)=z\sum_{i=0}^{w-1} q_i \bigg[\prod_{k=0, k\neq i}^{w-1}(e^{i E_k \tau}-z)\bigg]=z{\cal G}(z).
\end{equation}
Using $\sum_i q_i = 0$, it is not difficult to show that $\text{deg}({\cal D}(z))>\text{deg}({\cal G}(z)) $ (see details in Appendix \ref{order}).  We rewrite the generating function by ``general partial decomposition" for isolated poles of the denominator and a polynomial ${\cal G}(z)$ 
of order smaller than $w-1$.
Using the $w-1$ poles $\{ Z_i\}$ we found before, we rewrite ${\cal D}(z)=\beta(z-Z_1)(z-Z_2)\cdots(z-Z_{w-1})$ 
($\beta$ is the coefficient of $z^{w-1}$, see Appendix \ref{order}). Then we obtain
\begin{equation}
    \frac{{\cal G}(z)}{\beta(z-Z_1)...(z-Z_{w-1})}=\sum_{i=1}^{w-1}\frac{C_i}{Z_i(z-Z_i)} ,
    \label{C_i 32}
\end{equation}
%We may of course rewrite the summand on the right hand side as $C_i^{\prime}/(z-Z_i)$, with $C_i^{\prime}=C_i/Z_i$. 
%But it turns out more useful to use the representation in Eq. (\ref{C_i 32}). 
where $C_i$ is given by
\begin{align}
    C_i & =\frac{Z_i}{2\pi i}\oint_{\gamma_i}\frac{{\cal G}(z)}{\beta(z-Z_1)\cdots(z-Z_{w-1})}d z \nonumber \\
     & =\frac{{\cal N}_t(Z_i)}{\beta}\prod_{k\neq i}\frac{1}{Z_i-Z_k}.
\end{align}
The contours $\gamma_i$ enclose only $Z_i$ but not $\{Z_k\}_{k\neq i}$. Since $Z_i$ is the pole of $[{\cal D}(z)]^{-1}$, 
we can rewrite the multiplication as $\beta^{-1}\prod_{k\neq i}(Z_i-Z_k)^{-1}=[\partial_z {\cal D}(z)]^{-1}|_{z=Z_i}$, hence
\begin{equation}
    C_i=\frac{{\cal N}_t(Z_i)}{\partial_z {\cal D}(z)|_{z=Z_i}}.
    \label{C_i general}
\end{equation}
This allows us to rewrite the generating function as $\hat{\phi}_t(z)=\sum_{i=1}^{w-1}z C_i/[Z_i(z-Z_i)]$,
where $\hat{\phi}_t(z)$ is decomposed into the summation of the $z C_i/[(z-Z_i)Z_i]$ in which there is only one pole in the denominator.
With Eq. (\ref{phi-n}) the first detection amplitude reads
 \begin{equation}
     \phi_n=\sum_{i=1}^{w-1}\frac{C_i}{2\pi i}\oint_{\gamma}\frac{z^{-n}}{Z_i(z-Z_i)}dz=-\sum_{i=1}^{w-1}C_i Z_i^{-n-1}.
     \label{phi n}
 \end{equation}
The probability of finding the quantum state $|\psi_d\rangle$ at the $n^{\rm th}$ attempt is $F_n=|\phi_n|^2$. 
Summing the geometric series the total detection probability $P_{\rm det}=\sum_{n=1}^{\infty}F_n$ is
\begin{equation}
     P_{det}=\sum_{i,j=1}^{w-1}\frac{C_i C_j^{\ast}}{(Z_i Z_j^{\ast}-1)Z_i Z_j^{\ast}}.
     \label{P det excat}
\end{equation}
As mentioned before, other methods for finding $P_{\rm det}$ were considered in Ref. \cite{Thiel2019}. For a finite system, it was shown 
%in Ref. \cite{Thiel2019} 
that $P_\text{det}$ is independent of the measurement interval $\tau$ except for the special resonant points 
in Eq. (\ref{excptional points}) where new degeneracy appears. In finite-dimensional Hilbert space, the total detection probability is $P_{\rm det}=1$ 
when all the energy levels have projection on the detected state and the back-folded spectrum is not degenerate.

If the total detection probability is one, the detection of the quantum state in an experiment is guaranteed. We can define the mean 
FDT time $\langle t \rangle=\langle n \rangle \tau$, where $\langle n \rangle$ is the mean of the number of detection attempts.
For convenience, we call $\langle n \rangle$ the mean of FDT time in the rest of the paper due to the simple 
relation between the $\langle t \rangle$ and $\langle n \rangle$. From $\langle n \rangle=\sum_{n=1}^{\infty}n| \phi_n | ^2$, 
together with Eq. (\ref{phi n}), we find
\begin{equation}
    \langle n \rangle=\sum_{i,j=1}^{w-1}\frac{C_i C_j^{\ast}}{(Z_i Z_j^{\ast}-1)^2}.
    \label{mean n excat}
\end{equation}
Eqs. (\ref{Force field},\ref{C_i general},\ref{mean n excat}) expose how the mean FDT time depends on the spectrum of $H$, the initial state $|\psi_{\rm in}\rangle$, 
the detected state $|\psi_{\rm d}\rangle$ and the sampling time $\tau$. Since in general the denominator of Eq. (\ref{mean n excat})
is vanishing when some $Z_k$ is approaching the unit circle, we may have some critical scenarios, where the $\langle n \rangle$ can be 
asymptotically computed by neglecting non-diverging terms in the formal formula Eq. (\ref{mean n excat}). This leads to simpler 
formulas but with more physical insights. We will investigate these cases in the following sections.

\section{Relation of the mean FDT time and the FDR variance
\label{The detected transition time and return fluctuations}}

There is a general relation between the mean FDT time $\langle n \rangle$ and the matrix $\{V_{i,j}\}$, describing the variance of the FDR problem. The relation is rather general, but becomes especially useful when both $\langle n \rangle$ and $V_r$ are large.

First we reformulate some of the main equations which we will use later. The variance of the FDR time is \cite{Gruenbaum2013}
\begin{equation}
    V_r=\langle n^2 \rangle_r -\langle n \rangle_r^2=\sum_{i,j=1}^{w-1}V_{i,j},
    \label{V Gruen}
\end{equation}
where $V_{i,j}=2/(Z_i Z_j^{\ast}-1)$. Also $P_{\rm det}$ can be written in terms of summations over matrix elements of $P_{i,j}$:
\begin{equation}
    P_{\rm det}=\sum_{i,j}P_{i,j},\qquad P_{i,j}=\frac{C_i C_j^{\ast}}{(Z_i Z_j^{\ast}-1)Z_i Z_j^{\ast}}.
\end{equation}
Using Eq. (\ref{mean n excat}), the matrices $P_{i,j}$ and $V_{i,j}$ give also the mean FDT time:
\begin{equation}
    \langle n \rangle=\frac{1}{2}\sum_{i,j=1}^{w-1}Z_i Z_j^{\ast}P_{i,j}V_{i,j}.
    \label{n and var}
\end{equation}
This equation relates the $\langle n \rangle$ and terms $V_{i,j}$ of the $V_r$, which indicates that the
fluctuations of the FDR time reveal the characteristics of the mean FDT time. Below we show cases where one element of the summation is dominating $V_r \sim V_{s,s}$ and $|Z_s|\rightarrow 1$ (subscript $s$ stands for single.), such that

\begin{equation}
    \langle n \rangle \sim \frac{P_{s,s}}{2} V_{s,s}\sim \frac{P_{s,s}}{2}V_r.
    \label{einstein1}
\end{equation}
This is similar to the Einstein relation in the sense that diffusivity (a measure of fluctuations) is related to
mobility (a measure of the average response). In the Sec. \ref{time and fluctuation relation} we will find the exact 
expression for the different scenario. 

After obtaining the general results Eqs. (\ref{mean n excat},\ref{n and var}), we will focus on the diverging mean FDT time, 
where the asymptotic $\langle n \rangle$ and its relation to $V_r$ are developed. Eq. (\ref{mean n excat}) 
implies a divergent mean FDT time when $|Z_s|\rightarrow 1$. Since $|Z_s|=1/|z_s|$, where $z_s$ is the stationary point on the 
electrostatics field, the question is whether a stationary point is close to the unit circle. Next
we will investigate three scenarios where $|Z_s|\rightarrow 1$, using the electrostatic picture. We distinguish them into the following
cases: 1) a weak charge scenario, 2) two charges merging picture, and finally 3) one big charge theory.

\section{Weak charge\label{Weak charge}}
\label{sect:weak}

\begin{figure*}
    \centering
    \includegraphics[width=1.7\columnwidth]{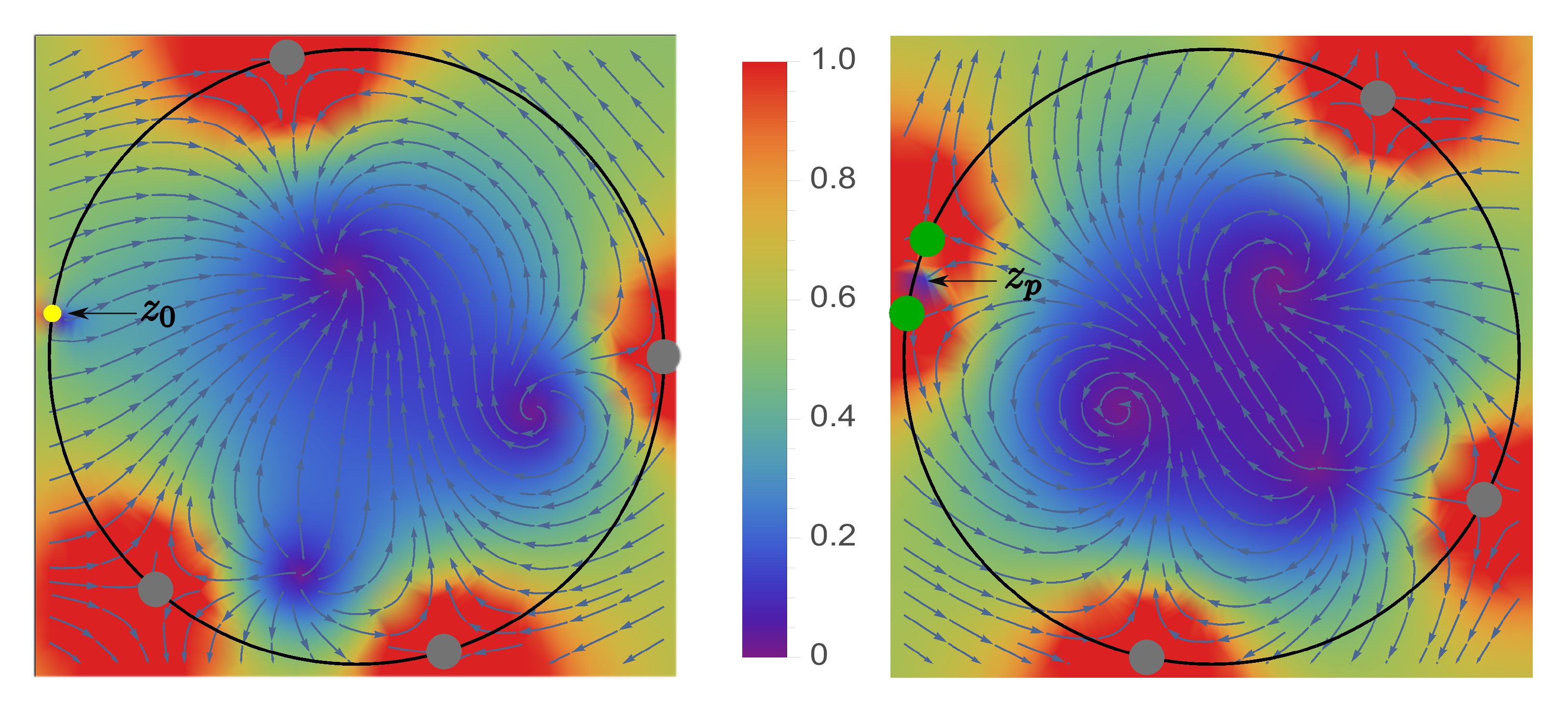}
    \caption{Schematic plot of the cases where the poles are close to the unit disk. From electrostatics if a charge (yellow)
    is weak, one stationary point denoted $z_0$ will be close to this weak charge, hence we have one pole $Z_0$ 
    that is close to the unit circle (left). Another case presented on the right is when two charges (green) are merging we also have one zero $z_p$ (or pole $|Z_p|=|z_p|^{-1}$ close to the unit circle. }
    \label{fig:poles for weak charge and two charges}
\end{figure*}

In electrostatics, when one charge becomes much smaller than all other charges, one of the stationary points will be close to the weak charge \cite{Gruenbaum2013} 
(see Fig. \ref{fig:poles for weak charge and two charges}, where the yellow charge indicates the weak charge, and its corresponding pole is $Z_0$). 
In analogy, the stationary point of the moon-earth system is much closer to the moon than to the earth. We denote this charge $p_0$ and the stationary
point $z_0$. The corresponding energy level of this weak charge is $E_0$ and its location is $\exp{(i E_0\tau)}$ on the unit circle. 
Since $z_0 \rightarrow e^{i E_0 \tau}$, from Eq. (\ref{relation of zero and pole}) the reciprocal pole $|Z_0|=1/|z_0|\rightarrow 1$. Using Eq. (\ref{mean n excat}), the asymptotic mean of the mean FDT time is
\begin{equation}
      \langle n \rangle \thicksim \frac{| C_0 |^2 }{(| Z_0|^2-1)^2},
      \label{weak 1}
\end{equation}
when $p_0\rightarrow 0$ and $|q_0|/p_0\gg1$.
Here we assume $|C_0|^2/(|Z_0|^2-1)$ is the dominating part of $\langle n \rangle$, and all other terms in 
Eq. (\ref{mean n excat}) are negligible. To find the exact expression of $\langle n \rangle$, we first need 
to find the pole $Z_0$. Using Eq. (\ref{Force field}) together with perturbation theory presented in the 
Appendix \ref{weak charge Appendix}, we get
\begin{equation}
     Z_0\thicksim e^{i\tau E_0}+\epsilon^{\ast}e^{2i \tau E_0},
     \label{Z_0}
\end{equation}
with
\begin{equation}
    \epsilon = \frac{p_0}{\sum_{k=1}^{w-1}p_k/(e^{i \tau E_0}-e^{i \tau E_k})}.
\end{equation}
Since $p_0\ll 1$, $e^{i E_0\tau}$ is the leading part of $Z_0$. Hence the pole $Z_0$ is located very close to 
the weak charge as we expect from basic electrostatics. The other $w-1$ charges give a small disturbance to $Z_0$ 
if they are not close to the weak charge. Substituting $Z_0$ into Eq. (\ref{C_i general}),
the coefficient $C_0$ reads (see Appendix \ref{weak charge Appendix}):
\begin{equation}
    C_0\sim -\frac{q_0}{p_0}\epsilon^{\ast}e^{2i\tau E_0}.
    \label{C_0}
\end{equation}
$C_0$ is determined by the fraction of $q_0$ and $p_0$, the parameter $\epsilon$ and the phase $e^{2i \tau E_0}$
which comes from the location of the weak charge. The small parameter epsilon is the effect of the remaining
charges in the system, excluding the weak charge, acting on a test charge, where the stationary point is found.

Finally, using the normalization condition $\sum_k p_k=1$ and $1/(1-\exp[i x])=1/2+i\cot{[x/2]}/2$, we get from Eq. (\ref{weak 1}) the mean FDT time
\begin{equation}
   \langle n \rangle\thicksim  \frac{|q_0|^2}{4p_0^2}\Bigg\{ 1+\bigg[\sum_{k=1}^{w-1} p_k \cot{[(E_k-E_0)\tau/2]}\bigg]^2\Bigg\} .
   \label{n weak}
\end{equation}
The prefactor $|q_0|^2/4p_0^2$ depends on $q_0$ and $p_0$ defined in Eqs. (\ref{qk},\ref{pk}), and they rely only on the stationary states with energy level $E_0$ the initial and final states, but not on the other energy states of the system. This prefactor is the envelope of the mean FDT time as the $\cot()$ solution is oscillating when we modify $\tau$. From our assumption 
$|q_0|/p_0\gg1$ the value of this envelope is large. The summation in the bracket shows that $\langle n \rangle$ depends 
on all charges as expected.

As mentioned when Eq. (\ref{excptional points}) holds we get the merging of two phases on the unit circle a case we will study in detail in the next section. In the vicinity of this point the mean FDT time diverges. So what is the physics for this divergence?
 We have shown before when two energy levels coalesce, the total detection probability $P_{\rm det}$ 
is not unity, which means the quantum particle goes to ``dark states" in the Hilbert space \cite{Thiel2019}. This divergence reflects that the total detection
probability $P_{\rm det}$ deviates from $1$, indicating that one or more states are not accessible by the quantum walker. We will see this 
connection in some examples below.

\section{Two merging charges\label{Two merging charges}}
\label{sect:merging}
Another case with a pole close to the unit circle is when the phases of two charges, denoted by $p_a$ and $p_b$, 
satisfy the resonance condition $\exp{(i E_a\tau)}\simeq \exp{(i E_b \tau)}$ (see Fig. \ref{fig:poles for weak charge and two charges}, the merging charges are colored green). As mentioned, this means that we are close to a degeneracy of the backfolded spectrum. 
It can be achieved by modifying $H$ or the sampling time $\tau$. Then the small parameter $\delta=(\Bar{E}_b-\Bar{E}_a)\tau/2$ 
measures the angular distance between the two phases. When the two charges merge, a related pole denoted $Z_p$ (subscript $p$ is for pair of merging charges), will approach the unit 
circle $|Z_p|\rightarrow 1$. Using Eq. (\ref{mean n excat}), 
for the mean FDT time $\langle n \rangle$, we get 
\begin{equation}
    \langle n \rangle \sim \frac{|C_p|^2}{(|Z_p|^2-1)^2},\quad \delta\rightarrow 0.
    \label{two merging charges}
\end{equation}
To find the pole $Z_p$, we first treat the charge field as a two-body system. Because by our assumption all other charges are far away 
from the two merging charges. Then we take the background charges into consideration. Using the two-body hypothesis together with 
Eq. (\ref{relation of zero and pole}), we find in perturbation theory (see Appendix \ref{two charge pole})
\begin{equation}
    Z_p\thicksim Z_p^{(0)}+Z_p^{(1)},
    \label{Z_p}
\end{equation}
here $Z^{(0)}_p$ and $Z^{(1)}_p$ are defined in Appendix \ref{two charge pole} in Eq. (\ref{appendix Z_p}). Plugging $Z_p$ into Eq. (\ref{C_i general}) yields for the coefficient $C_p$
\begin{equation}
    |C_p|\sim 2\delta\frac{|q_a p_b -q_b p_a|}{(p_a +p_b )^2} ,
    \label{C_p}
\end{equation}
where $|C_p|$ is determined by the phase difference, charges and $q_k$. Since $\delta$ is a small parameter, 
$|C_p|$ also becomes small when two charges merge. Substituting $C_p$ and $Z_p$ into Eq. (\ref{two merging charges}), the mean FDT time
becomes
\begin{equation}
    \langle n \rangle \thicksim \frac{(p_a+p_b)^2|q_a p_b-q_b p_a|^2}{p_a^2p_b^2}\frac{1}{\tau^2(\Bar{E}_b-\Bar{E}_a)^2} .
    \label{n two charges}
\end{equation}
It should be noted that this formula does not include the background, which is quite different from the weak charge case. When two charges are merging, the expected transition time $\langle n \rangle$ diverges since $(\Bar{E}_b-\Bar{E}_a)^2\tau^2$ is small.
The term $|q_a p_b-q_b p_a|^2$ comes from the interference. At the special case $|q_a p_b-q_b p_a|^2=0$ we have an elimination of the resonance, meaning that the effect of divergence might be suppresses.

\section{Relation between mean FDT time and FDR fluctuations}
\label{time and fluctuation relation}

When there is only one pole dominating, simple relations between the mean FDT time and the fluctuations of the FDR time are found. 
We start from the general relation Eq. (\ref{n and var}). When the pole $|Z_s|\rightarrow 1$ we have  Eq. (\ref{einstein1}). Here $Z_s$ is a single pole approaching the unit circle, it could be either $Z_p$ for two merging charges or $Z_0$ for one weak charge.
$P_{s,s}$ is the diagonal term of the matrix $\{P_{i,j}\}$, which is real and positive. 
Based on Secs. \ref{Weak charge},\ref{Two merging charges} we can get exact expressions for $P_{s,s}$ under different circumstances.

In the weak charge regime, $P_{s,s}\sim |C_0|^2/(|Z_0|^2-1)$. Substituting the $C_0$ and $Z_0$ into $P_{s,s}$, the ratio of the mean FDT time 
and the FDR variance reads
\begin{equation}
   \frac{\langle n \rangle}{V_r} \sim \frac{|q_0|^2}{2p_0},
   \label{n v weak q}
\end{equation}
when energy level $E_0$ is not degenerate, we have:
\begin{equation}
    \frac{\langle n \rangle}{V_r} \sim \frac{|\langle \psi_{\rm in}|E_0\rangle|^2}{2}.
    \label{n v weak}
\end{equation}
From Eq. (\ref{n v weak q}) and Eq. (\ref{n weak}), we can get the expression of $V_r$, which confirms the result for $V_r$ in \cite{Yin2019}. 
The beauty of this simple relation is that it only depends on the overlap of the initial state $|\psi_{\rm in}\rangle$ and $|E_0\rangle$. 
So how we prepare the quantum particle is of great importance for the mean FDT time. The quantum particle will remember its history. Furthermore, $ |\langle \psi_{\rm in}| E_0\rangle|^2/2<1/2$
implies that the mean FDT time is bounded by one half of the FDR variance.

For the two merging charges scenario we have $P_{s,s}\sim |C_p|^2/(|Z_p|^2-1)$. Using Eqs. (\ref{Z_p},\ref{C_p}) gives us the ratio
\begin{equation}
    \frac{ \langle n \rangle}{V_r}\sim\frac{|q_a p_b-q_b p_a|^2}{2(p_a+p_b)p_a p_b} .
    \label{n v two charge}
\end{equation}
From Eqs. (\ref{n two charges},\ref{n v two charge}) we get an expression for $V_r$, which was also derived in \cite{Yin2019}. 
Here the initial state $|\psi_{\rm in}\rangle$ plays an important role because $q_a$ and $q_b$ are related to the initial state (unlike $p_a$ and $p_b$).
Under some special symmetry of the system we can get $p_a/q_a=p_b/q_b$, such that $|q_a p_b-q_b p_a|^2=0$. As mentioned,
this is reflects a elimination of the resonance because $\langle n \rangle$ will tend to some small values, while the mean FDR variance diverges.

\textbf{Remark:} We may start from Eqs. (\ref{mean n excat},\ref{V Gruen}), if one of the poles is denoted $Z_s$ and is close
to the unit circle. Then we have roughly $\langle n \rangle \sim |C_s|^2/(|Z_s|^2-1)^2$ and $V_r\sim 2/(|Z_s|^2-1)$. The relation of the mean FDT time and 
the FDR variance is $\langle n \rangle \sim |C_s|^2V_r^2/4$, i.e., $\langle n\rangle$ is proportional to $V_r^2$. 
This intuition does not reveal the real physics, since for a divergent $V_r$ we get $|C_s|\rightarrow 0$.

\subsection{Two-level System\label{Two level system}}

\begin{figure}
    \centering
    \includegraphics[width=1\columnwidth]{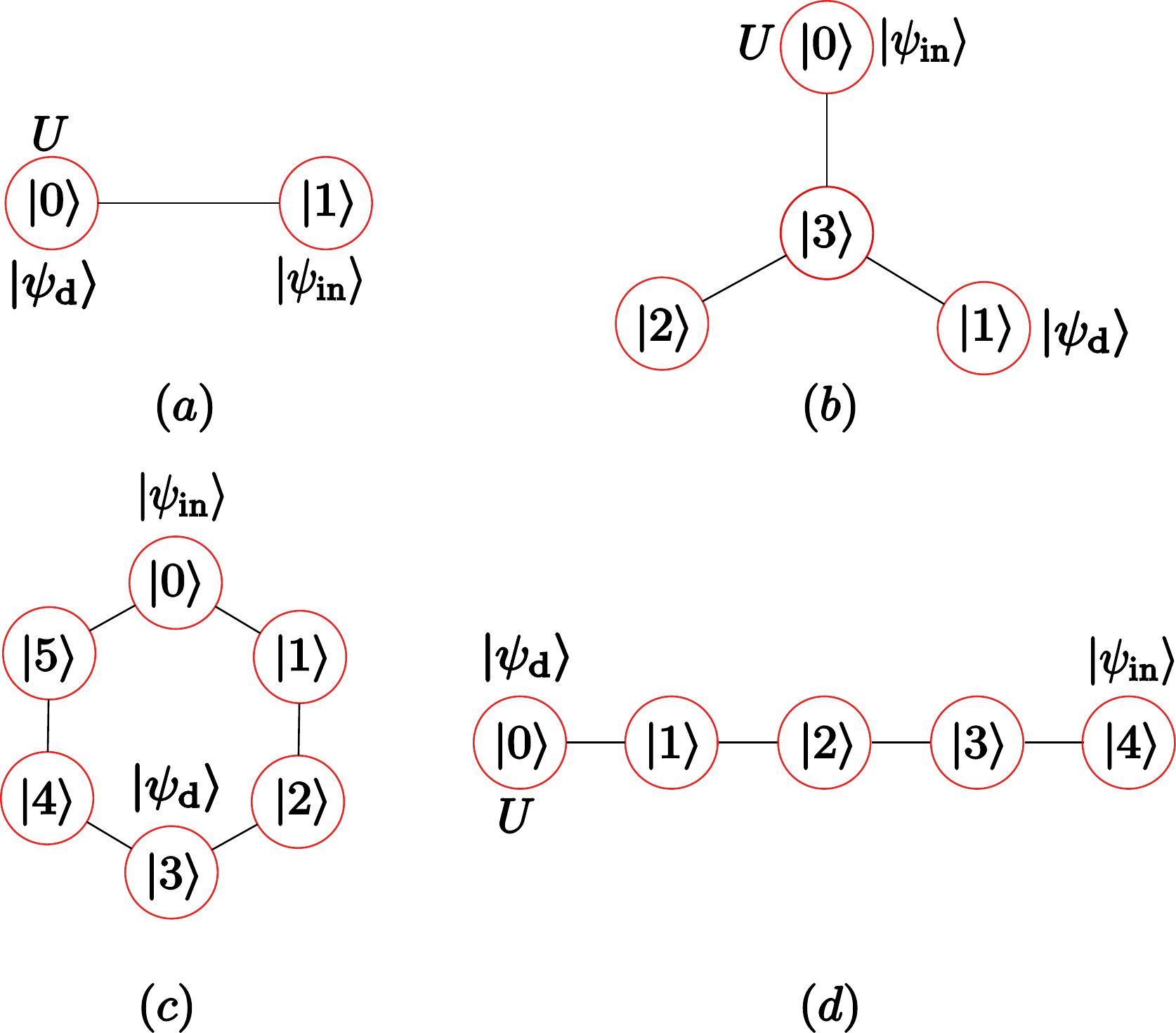}
    \caption{Schematic models. We perform the calculations on different graphs. The quantum particle is prepared in the initial 
    state $| \psi_{\rm in} \rangle$ and we set the detector to detect the state $|\psi_{\rm d}\rangle$. 
    $U$ is the strength of potential well or potential barrier we set in the system. $(a)$ Two level model. $(b)$ Y-shaped molecule. 
    $(c)$ Benzene-like ring. $(d)$ Linear five-site molecule. }
    \label{fig:schematic model}
\end{figure}

As an application of our general theory we consider tight-binding models on simple graphs. 
The first example is a quantum walk on a two-site graph (see Fig. \ref{fig:schematic model}($a$)) (i.e. a two-level system).
The Hamiltonian of this system reads
\begin{equation}
    H=-\gamma(| 0 \rangle\langle 1|+| 1\rangle\langle 0|+U| 0\rangle\langle 0|).
    \label{two level H}
\end{equation}
It describes a quantum particle hopping between two sites 0 and 1, where a potential $U$ is added at site 0. 
This model also presents a spin $1/2$ in a field.

We prepare the initial quantum state as $|0\rangle$, which means that the particle is on site 0. 
The detector is set to detect the particle at site 1; i.e. the detected state is $|1\rangle$. 
From Eq. (\ref{two level H}) the energy spectrum of the system is (we set $\gamma=1$ subsequently): 
$E_0 = (-U-\sqrt{U^2+4})/2$ and $E_1 = (-U+\sqrt{U^2+4})/2$. In the large $U$ limit, where $E_0\rightarrow -U$ and $E_1\rightarrow 0$, 
the two energy levels $E_0$ and $E_1$ are separated. From Eq. (\ref{pk}) the charge $p_0=1/(E_0^2+1)$ and from normalization $p_1=1-p_0$.
When we increase the value of the potential $U$, the charge $p_0\rightarrow 0$, which represents a weak charge in the system. 
From Eq. (\ref{qk}) we have $q_0=E_0/(E_0^2+1)$ and $q_1=E_1/(E_1^2+1)$. The ratio $|q_0|/p_0$ is $|E_0|$ ,which is our dimensionless variable growing with the potential $U$. From Eq. (\ref{n weak}) the mean FDT time of this simple two-level system is
\begin{equation}
    \langle n \rangle \sim \frac{U^2}{4}[ 1+\cot^2{(U\tau/2)}].
    \label{n two-level}
\end{equation}
$\langle n \rangle$ becomes larger as we increase $U$, indicating the potential well blocks the propagation of the wave function, 
making it hard to find the particle at the detected state. In Eq. (\ref{n two-level}), when $U\tau$ is close to $2\pi k, k=1,2,\cdots$ 
the mean FDT time diverges. Note that $U \tau = 2 \pi k$ is the condition for exceptional points (Eq. (\ref{excptional points})), in the limit of large $U$.  At these exceptional points, the total detection probability $P_{det}$ drops from $1$ to $0$.

\begin{figure}
    \centering
    \includegraphics[width=0.99\columnwidth]{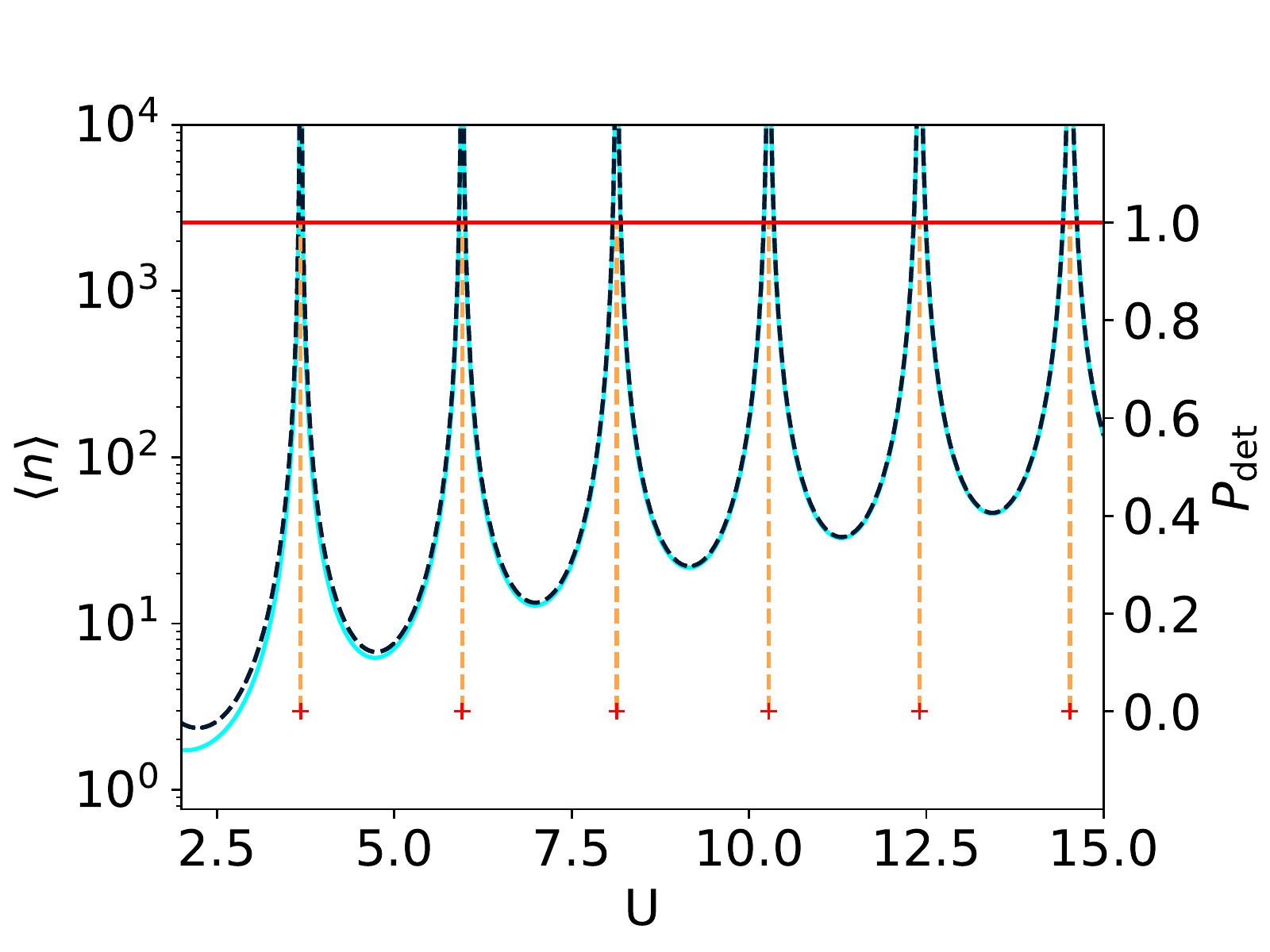}
    \caption{The mean FDT time $\langle n \rangle$ (left y axis) and total detection probability $P_{\rm det}$ (right y axis) versus the potential $U$ 
    of the two level system for the transition from $|1\rangle\rightarrow|0\rangle$ in Fig. \ref{fig:schematic model}($a$). Here we fix $\tau=3$. 
    The exact mean FDT time (black dashed line) meets quite well with our theoretical result Eq. (\ref{n two-level}) (cyan line). Close to the exceptional 
    points $U\sim 2\pi k/\tau, k=1,2,\cdots$, where the back-fold energy levels are degenerate, the total detection probability (red line) drops to $P_{\rm det}=0$
    and $\langle n \rangle$ diverges as we expected.}
    \label{fig:two level system}
\end{figure}

Choosing the sampling frequency $1/\tau=1/3$, the exact $\langle n \rangle$ can be obtained either from the quantum renewal equation Eq. (\ref{quantum renewal})
or our first main result Eq. (\ref{mean n excat}). Here we use the latter formula, and the result is visualized in Fig. \ref{fig:two level system} (left y axis).
In the vicinity of the exceptional points the total detection probability drops from the unity and the mean FDT time diverges.

\subsection{Y-shaped Molecule}

\begin{figure*}
    \centering
    \includegraphics[width=1.99\columnwidth]{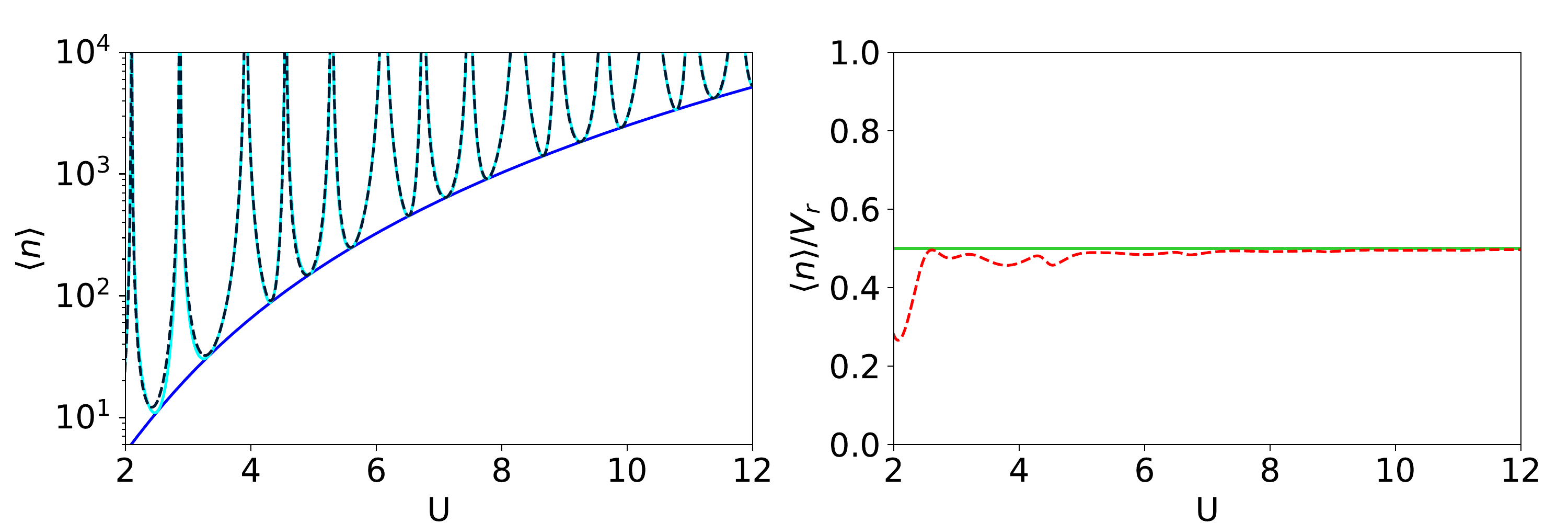}
    \caption{The mean FDT time $\langle n \rangle$ (left) and the ratio of the mean FDT time and the FDR variance $\langle n \rangle/V_r$ (right) versus $U$ for the Y-shaped model. 
    Here we choose $\tau=3$. The quantum particle travels from $|0\rangle$ to $|1\rangle$. The $\langle n \rangle$ diverges when $U$ is close to the exceptional points. The weak charge envelope (blue line) clearly gives the tendency of the transition time. 
    The theoretical result Eq. (\ref{n Y shape}) (cyan line) fits quite well with the exact $\langle n \rangle$ (black dished line). For $\langle n \rangle/V_r$, the exact result 
    (red dished line) gradually approaches to our theoretical value (green solid line), while $\langle n \rangle/V_r\leqslant 1/2$ in the whole regime. The fluctuations of the FDR give the upper bound 
    of the corresponding mean FDT time in this example.}
    \label{fig:Y-shaped n n/v}
\end{figure*}

The next example is the Y-shaped molecule, where the quantum particle can jump from states $|0\rangle, \quad |1\rangle, \quad |2\rangle$ to state $|3\rangle$ and vice versa (see schematics in Fig. \ref{fig:schematic model} ($b$)). 
We add a potential $U$ at site 0. Then the Hamiltonian of the Y-shaped molecule reads 
\begin{equation}
    H=-\gamma(U|0\rangle\langle0|+\sum_{i=1}^3|3\rangle\langle i|+\sum_{j=1}^3|j\rangle\langle 3|).
\end{equation}
We prepare the quantum particle in the state $|\psi_{\rm in}\rangle=|0\rangle$ and the detection is performed in the state $|1\rangle$. Due to the mirror 
symmetry of Y-shaped molecule, the energy level $E_3=0$. Other energy levels $E_0$, $E_1$ and $E_2$ are given by the roots of the equation
$E^3+U E^2 - 3E-2U = 0$. When $U$ is large, we have $E_0\sim -U$, $E_1\sim \sqrt{2}$ and $E_2\sim -\sqrt{2}$.
 From Eq. (\ref{pk}) the charges are $p_0\rightarrow 0$, $p_1\rightarrow 1/4$, $p_2\rightarrow 1/4$ and $p_3\rightarrow 1/2$. The appearance of the weak charge $p_0$ is because one of the eigenstate is nearly localised on $|0\rangle$, more specifically $|E_0\rangle \simeq |0\rangle$. The exact numerical values of both energy levels $\{E_i\}$ and charges $\{p_i\}$ are shown in Appendix \ref{order}
in Fig. \ref{fig: energy level charge dnesity}. Using Eq. (\ref{n weak}) the mean FDT time of the Y-shaped molecule reads
\begin{equation}
    \langle n \rangle \sim \frac{|q_0|^2}{4 p_0^2}\Bigg\{1+\bigg[\sum_{i=1}^3 p_i\cot{[(E_i-E_0)\tau/2]}\bigg]^2\Bigg\}.
    \label{n Y shape}
\end{equation}
The initial site and detected site are not symmetric because of the potential $U$. This implies
$|\langle E_0|0\rangle|\gg |\langle E_0| 1\rangle|$ and $|q_0|/p_0\gg 1$. 
When two energy levels are coalescing Eq. (\ref{n Y shape}) diverges. The prefactor in Eq. (\ref{n Y shape}) 
indicates the asymptotic tendency of the mean FDT time versus the potential $U$ (see Fig. \ref{fig:Y-shaped n n/v}), which should be observed experimentally. 
We denote this prefactor as the weak charge envelope $\langle n \rangle_e$ 
\begin{equation}
    \langle n \rangle_{e}\sim \frac{|q_0|^2}{4 p_0^2}= \frac{1}{4}\frac{|\langle 0| E_0\rangle|^2}{|\langle 1| E_0\rangle|^2}.
    \label{n Y e}
\end{equation}
The weak charge envelope is determined by the overlaps of the initial and detected state. From Eq. (\ref{n v weak}) the relation between 
the mean FDT time and the FDR variance gives 
\begin{equation}
    \frac{\langle n \rangle}{V} \sim \frac{1}{2}.
    \label{n v Y-shaped}
\end{equation}

To plot an example, we solve the quantum renewal equations exactly, as was done in Sec. \ref{Two level system}, here we choose the sampling period $\tau=3$. 
The value of potential well $U$ goes from $2$ to $12$. As shown in Fig. \ref{fig:Y-shaped n n/v}, Eqs. (\ref{n Y shape}, \ref{n Y e}, \ref{n v Y-shaped})
work well in the weak charge regime where $U$ is large.

\subsection{Benzene-type ring}\label{Benzene-type ring}
\label{sect:ring}

For the third model we consider the Benzene-type ring which has six spacial states $|0\rangle,|1\rangle,\cdots,|5\rangle$ (see Fig. \ref{fig:schematic model}($c$)). 
We use periodic boundary conditions and thus from the site labeled $x=5$ the particle may hop either to the origin $x=0$ or to the site labeled $x=4$.
Then the Hamiltonian of the ring reads
\begin{equation}
    H=-\gamma[\sum_{x=0}^5(| x \rangle\langle x+1|+| x+1\rangle\langle x|)].
\end{equation}
We prepare our quantum particle in the state $|0\rangle$ and perform the detection in the state $|3\rangle$,
which monitors the travel of the quantum particle from site $0$ to the opposing site. In this case $P_{\rm det}=1$ except for special sampling times.

\begin{figure*}
    \centering
    \includegraphics[width=1.8\columnwidth]{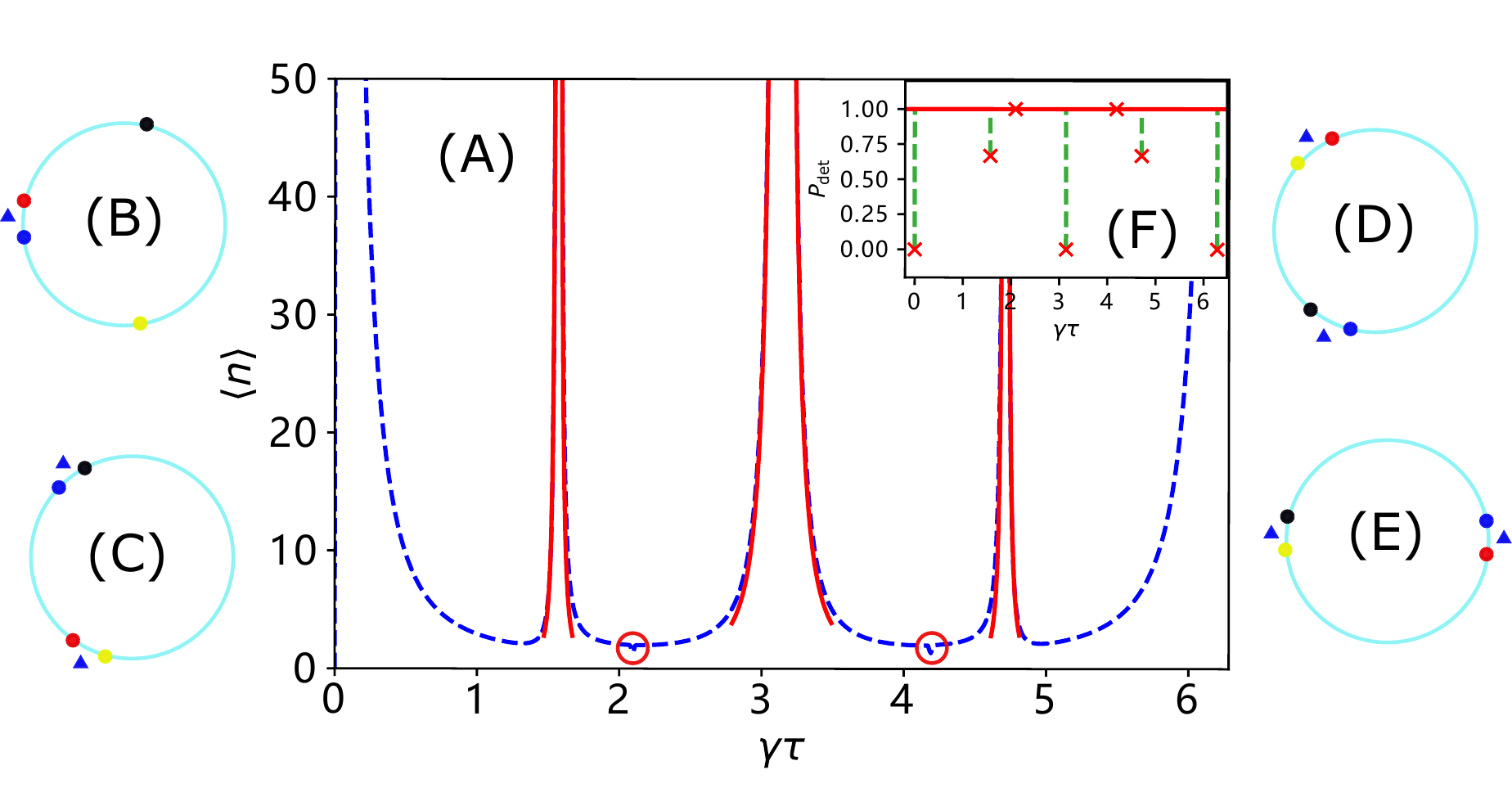}
    \caption{The mean FDT time $\langle n \rangle$ versus $\gamma\tau$ of the Benzene-type ring.  The quantum particle is prepared at $|0\rangle$ and the detector is set on the opposite site (see Fig. \ref{fig:schematic model}($c$)). When $\tau\rightarrow \pi/2$ or $3\pi/2$, the charges $p_1$ (blue) and $p_2$ (red) come close to each other (B). The $P_{\rm det}$ (F) drops to $2/3$, so both the two-charge theory  
    Eq. (\ref{n ring t=pi/2}) (red curve) and the exact result (blue curve) diverge. When $\tau\rightarrow 2\pi/3$ or $4\pi/3$, there are two groups of charges merging (C) and (D). 
    However because of the elimination of the resonance the leading term in the mean FDT time $\langle n \rangle$ vanishes (see Eqs. (\ref{n two charges},\ref{Bring 0})) and it jumps to some small value instead of diverging as usual. 
    The total detection probability remains unity. In the graph we have the ``discontinuity" close to these points. The blue point represents the charge $p_1$, the red is $p_2$, the yellow 
    is $p_3$ and the black is $p_4$. (B) $\tau\rightarrow \pi/2$ or $3\pi/2$. (C) $\tau\rightarrow 2\pi/3$. (D) $\tau\rightarrow 4\pi/3$. (E)$\tau\rightarrow \pi$.}
    \label{fig:ring lacation n pure}
\end{figure*}

The Hamiltonian of the benzene-type ring has the energy spectrum $E_k=-2\cos{(\theta_k)}$ and the eigenstates are 
$|E_k\rangle^T =(1,e^{i \theta_k},e^{2i \theta_k},e^{3i \theta_k},e^{4i \theta_k},e^{5i \theta_k})/\sqrt{6}$ 
with $\theta_k=2\pi k/6$ and $k=0,1,2,3,4,5$  (the superscript $T$ is the transpose). In this case we have four distinct energy levels so $w=4$.
From Eqs. (\ref{qk},\ref{pk}) the charges and $q_k$ read
\begin{equation*}
    \begin{aligned}
        p_1 &=&\frac{1}{6},\quad p_2 &=& \frac{1}{6},\quad p_3 &=& \frac{1}{3}, \quad p_4 &=& \frac{1}{3};\\
     \quad q_1 &=& \frac{1}{6},\quad q_2 &=& -\frac{1}{6},\quad q_3 &=& -\frac{1}{3}, \quad q_4 &=& \frac{1}{3}.
    \end{aligned}
\end{equation*}
As mentioned, The energy spectrum of the ring is degenerate and the sampling time $\tau$ will introduce effective degeneracies to the problem. 
From Eq. (\ref{excptional points}), the exceptional sampling times are $\tau=\pi/2,2\pi/3,\pi,4\pi/3, 2\pi$ in the time interval 
$\{\tau | 0\leqslant \tau\leqslant2\pi\}$. Close to these exceptional points we will have the scenario of two charges merging, 
where we can employ our equations to give the theoretical predictions (see Fig. \ref{fig:ring lacation n pure}).
\begin{itemize}
    \item[1.]
    When $\tau$ is close to $\pi/2$ or $3\pi/2$ we have $|E_1-E_2|\tau\sim 2\pi k$. The charges $p_1$ and $p_2$ coalesce (see Fig. \ref{fig:ring lacation n pure}(B)). 
    For the mean transition time $\langle n \rangle$ and $\langle n \rangle/V_r$, using Eqs. (\ref{n two charges},\ref{n v two charge}) we have
    \begin{equation}
        \langle n \rangle\sim\frac{1}{36}\frac{1}{(\tau-\pi/2)^2},\quad \frac{\langle n \rangle}{V_r}\sim\frac{1}{6}.
        \label{n ring t=pi/2}
    \end{equation}
    \item[2.]
    When $\tau$ is close to the $2\pi/3$ or $4\pi/3$ we have $|E_1-E_4|\tau\sim 2\pi k$ and $|E_2-E_3|\tau\sim 2\pi k$. 
    Two pairs of charges are merging separately (see Fig. \ref{fig:ring lacation n pure}(C) and (D)). From Eqs. (\ref{C_p},\ref{n two charges}), due to the elimination 
    $q_1 p_4-q_4 p_1=0$ and $q_2 p_3-q_3 p_2=0$ we have
    \begin{equation}
        \langle n \rangle\sim O(1), \quad \langle n \rangle/V_r\rightarrow 0 .% V= 0.
        \label{Bring 0}
    \end{equation}
    The leading order of $\langle n \rangle$ vanishes, so $\langle n \rangle$ drops to some small values, leading to a small ``discontinuity" on the graph. 
    Close to these points we find that it takes less time for the walker to reach the detected state.
    \item[3.]
    When $\tau$ is close to $\pi$ we also have two groups of charges merging Fig. \ref{fig:ring lacation n pure}(E), i.e. $p_1$ is close to $p_2$, and $p_3$ is close to $p_4$. 
    Eq. \ref{n two charges} gives
    \begin{equation}
        \langle n \rangle\sim \frac{1}{36}\frac{1}{(\tau-\pi)^2}+\frac{4}{9}\frac{1}{(\tau-\pi)^2}.
        \label{ring 3}
    \end{equation}
     For the ratio of $\langle n \rangle$ and $V_r$ there are two groups of charges which we treat separately. 
     The first group we use Eq. (\ref{n v two charge}) to obtain $\langle n \rangle_{1,2} /V_{1,2}=1/6$. 
     Similarly, for the second group we have $\langle n \rangle_{3,4} /V_{3,4}=1/3$. The return variance $V_r=V_{1,2}+V_{3,4}$ and the mean FDT time $\langle n \rangle=\langle n \rangle_{1,2}+\langle n \rangle_{3,4}=V_{1,2}/6+V_{3,4}/3$. We can measure the fluctuations $V_r$ but not the terms $V_{1,2}$ and $V_{3,4}$. So here we do not have a direct relation between $\langle n \rangle$ and $V_r$. Using Eqs. (\ref{two merging charges},\ref{n v two charge}), we first calculate $V_{1,2}$ and $V_{3,4}$ (then $V_r=V_{1,2}+V_{3,4})$. Comparing $V_r$ and Eq. (\ref{ring 3}), we have $\langle n \rangle/V_r=5/18$. 
     
\end{itemize}

As shown in Fig. \ref{fig:ring lacation n pure}, we plot the exact results of the $\langle n \rangle$ for the $\tau$ from $0$ to $2\pi$. 
The theoretical predictions meet the exact values quite well close to the exceptional points where the total detection probability exhibits a sudden jump in its value.

So far we deal with one zero close to unit circle,
and now we switch to the more complicated cases where we have
more than one pole in the vicinity of the unit circle. We find the mean FDT time, but an Einstein like relation is not achieved in such case (as an example, see part 3 of the Benzene-type ring).

\section{Big charge theory\label{Big charge theory}}
\label{sect:big}

\begin{figure}
    \centering
    \includegraphics[width=0.9\columnwidth]{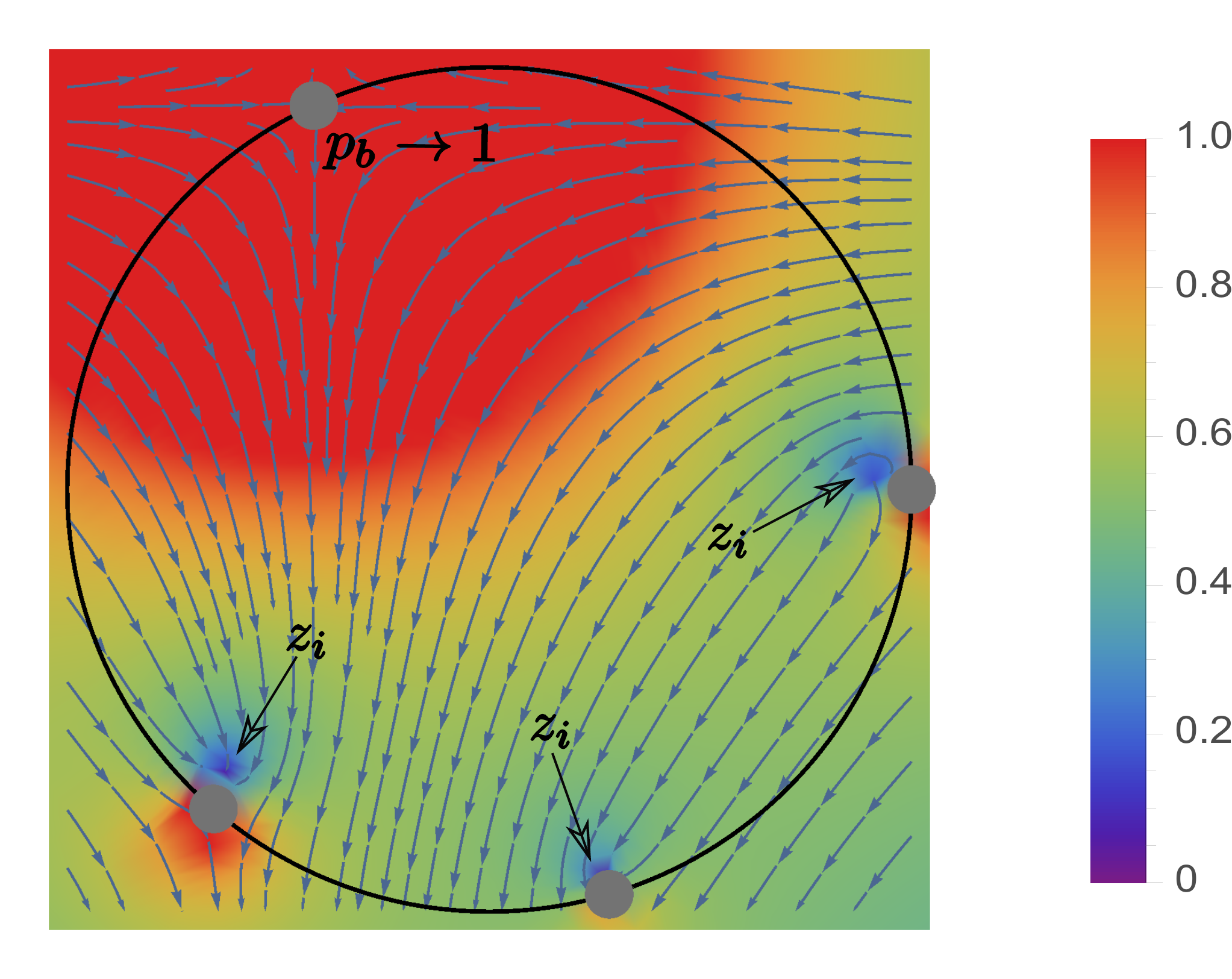}
    \caption{Schematic plot of the zeros $\{z_i\}$ in the complex plane for the big charge theory. Here $p_b\rightarrow 1$ is the big charge and hence from normalization all other charges are weak. The stationary points ($\{z_i\}$) are close to the weak charges, so as mentioned in the main text the poles $|Z_i|=1/|z_i|\rightarrow 1$, and they are all out side the unit circle.}
    \label{fig:big charge charge poles}
\end{figure}
Another scenario which leads to divergences of the mean FDT time $\langle n \rangle$ is when all the poles are close to the unit circle. This comes from the fact that the detected state is close 
to one of the eigenstates of Hamiltonian $H$, leading to a big charge appearing in the theory (Eq. (\ref{pk}). Using Eq. (\ref{mean n excat}),
the off-diagonal terms in $\langle n \rangle$ are negligible compared with the diagonal terms, then we get
\begin{equation}
    \langle n \rangle\sim\sum_{i=0,i\neq b}^{w-1}\frac{|C_i|^2}{(-1+|Z_i|^2)^2}\ ,\ \ |Z_i|\sim 1.
\end{equation}
The big charge, denoted $p_b\sim 1$, associated to the energy level $E_b$, is large in comparison with the other charges. 
Since the sum of all the charges is unity $\sum_{k=1}^{w-1}p_k=1$ and each of them is positive we have $1-p_b=\sum_{k\neq b}p_k\sim0$. 
Hence there is one big charge $p_b$ and $w-1$ weak charges. Basic electrostatics indicates that the $w-1$ stationary points will lie close to the $w-1$ weak charges. 
From Eq. (\ref{relation of zero and pole}) we have $|Z_i|=1/|z_i|$, such that all the poles $|Z_i|\rightarrow 1$ in this case. 
As visualized in Fig. \ref{fig:big charge charge poles}, the $w$ charges have $w-1$ poles and all of them are close to the weak charges.

Because all the charges are weak except for $p_b$, we find a stationary point $z_i$ when we consider only a pair of charges, i.e. $p_b$ and one of the $w-1$ weak charges $p_i$. 
This problem becomes a two-body problem (the charge $p_b$ and the weak charge $p_i$) for finding the stationary point between them, and all other charges are negligible. Using Eq. (\ref{Force field}), the zeros are given by the root 
of $p_b/(e^{i E_b \tau}-z_i)+p_i/(e^{i E_i \tau}-z_i)=0$, which yields $z_i=(p_i e^{i E_b \tau}+p_b e^{i E_i \tau})/(p_i+p_b)$.
From the relation between zeros and poles in Eq. (\ref{relation of zero and pole}) we have
\begin{equation}
    Z_i\sim e^{i E_i \tau}+(e^{i E_i \tau}-e^{i( 2E_i-E_b) \tau})\frac{p_i}{p_b}.
\end{equation}
The $i$ goes from $i=0$ to $w-1$ but $i\neq b$, so all $w-1$ poles are found. The first part of $Z_i$ is just the location of the charge $p_i$, 
the second part is small and comes from the net field of $p_i$ and $p_b$. We put the $Z_i$ into Eq. (\ref{C_i general}) to get the coefficient
\begin{equation}
    C_i\sim -\frac{q_i}{p_b}(1-e^{i (E_i-E_b) \tau})e^{i E_i \tau}.
    \label{big Ci}
\end{equation}
Here enters the ratio of $q_i$ and the big charge $q_i/p_b$,  which is a small parameter. $e^{i (E_i-E_b)}$ measures the phase difference between them. Substituting both $Z_i$ and $C_i$ into Eq. \ref{C_i general}, the mean FDT time for the 
big charge scenario reads
\begin{equation}
    \langle n \rangle\sim \sum_{i=0,i\neq b}^{w-1}\frac{|q_i|^2}{4p_i^2\sin^2{[(E_i-E_b)\tau/2]}} .
    \label{n big charge}
\end{equation}
It is very interesting to recall that in our weak charge theory (see Eq. (\ref{n weak})) the envelope is given by $|q_0|^2/4p_0^2$, where $p_0$ is a weak charge. 
For the big charge we have $|q_i|^2/4p_i^2$, where $p_i$ is also small.

\subsection{Localized wave function}

A good example for the big charge theory is when the wave function is effectively localized at the detected state by a strong potential. 
For instance, we localize the wave function at one node of the graph, and then we set our detector at this node.
To establish a specific example, we choose a five-site linear molecule put the detector at the site $x=0$ and prepare the initial state at $|4\rangle$. In order to localize the wave function 
at the detected state, we add a potential barrier $U$ at site $x=0$ (see Fig. \ref{fig:schematic model}($d$)). Then the Hamiltonian of this five-site molecule reads
\begin{equation}
    H=-\gamma[\sum_{x=0}^4 (| x \rangle\langle x+1|+| x+1\rangle\langle x|)+U| 0\rangle\langle 0|].
    \label{H five}
\end{equation}
Here the boundary conditions are that from the site labeled $x=4$ one can only hop to the site labeled $x=3$, and from the site labeled $x=0$ 
one can only hop to the site labeled $x=1$.   

For the energy spectrum we consider the regime where the wave function is effectively localized. As we increase the value of $U$, the energy level $E_0\rightarrow -U$. At the same time, this large potential well makes it difficult for the quantum particle 
to hop to the state $|0\rangle$. So the remaining four energy levels are given by the new Hamiltonian 
$H_l=-\gamma\sum_{x=1}^4(| x \rangle\langle x+1|+| x+1 \rangle\langle x|)$ with the same boundary condition as Eq. (\ref{H five}). Hence the energy spectrum reads
$E_0 \sim -U$, $E_1 \sim (1+\sqrt{5})/2$, $E_ 2\sim -(1+\sqrt{5})/2$, $E_3 \sim (-1+\sqrt{5})/2$ and $E_4 \sim (1-\sqrt{5})/2$. Notice that the energy levels are non-degenerate hence $w=5$.
The exact values of the energy levels are calculated and depicted in Appendix \ref{order} in Fig. \ref{fig: energy level charge dnesity} $(C)$.

Next we prepare the quantum particle in the state $|4\rangle$, such that the system describes the movement of the particle from site $x=4$ to $x=0$ on a 
linear molecule. From Eq. (\ref{pk}) it follows that the big charge $p_0\rightarrow 1$ and the remaining weak charges $p_{i\neq 0}\rightarrow 0$. 
With Eq. (\ref{n big charge}) we get for the mean FDT time 
\begin{equation}
    \langle n \rangle\sim \sum_{i=1}^4\frac{|q_i|^2}{4p_i^2\sin^2{[(E_i-E_0)\tau/2]}}.
    \label{n 5 sites}
\end{equation}

\begin{figure}
    \centering
    \includegraphics[width=0.99\columnwidth]{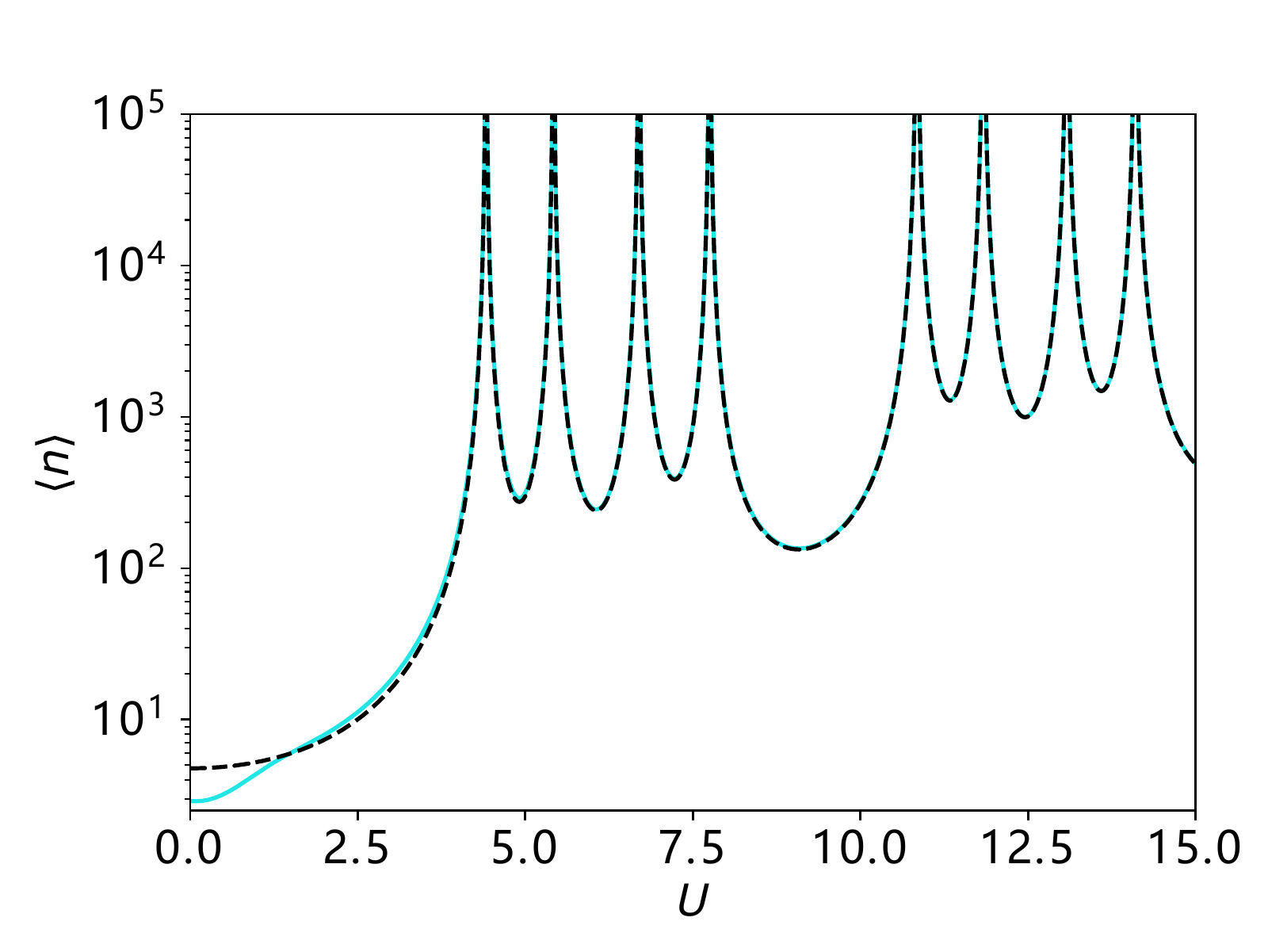}
    \caption{The mean FDT time $\langle n \rangle$ versus $U$ for the five-site molecule. Here the sampling time is $\tau=1$ and the quantum particle moves
    from $|4\rangle$ to $|0\rangle$, see Fig \ref{fig:schematic model} ($d$). The exact values (black dashed line) are calculated from Eq. (\ref{mean n excat}). 
    The cyan line is our big charge theory result Eq. (\ref{n 5 sites}). Close to the exceptional points given by Eq. (\ref{excptional points}), the mean FDT times diverge. 
    Since the strong charge $p_0$ rotates two laps on the unite circle, there are two clusters of peaks. In each cluster the big charge passes through the remaining four charges, leading to the four peaks 
    in the graph.}
    \label{fig:five sites n t n}
\end{figure}

In Fig. \ref{fig:five sites n t n} we compare the numerical result with our big charge theory, choosing the sampling time $\tau=1$ and the potential well from $0$ to $15$. 
%Each time when the big charge $p_0$ meets other charges, the mean FDT time diverges. Since the energy level of %$p_0$ increases with $U$  and remaining
%energy levels are nearly constant, hence there are two groups of diverging peaks and in each group there are four %peaks, corresponding to the four charges.  
In the limit of large $U$ the four weak charges
are fixed on the unit circle, their positions are given by their $U$ independent phase $\exp{(i E_i\tau)}$. 
When we increase $U$ the strong charge $p_0$, is thus crossing the location of the other charges and in the range $0<U<15$ which happens twice ($15/2\pi \sim 2$). As shown in Fig. \ref{fig:five sites n t n}, we have two groups of divergencies each with four peaks. The number of peaks in each group is $w-1=4$.

\section{discussion}\label{discussion}
We have used the quantum renewal equation \cite{Friedman2017} to investigate the mean FDT time, for systems in a finite-dimensional Hilbert space. A general formula for mean FDT time is developed. Then we focus on the diverging mean FDT times and find a relation similar as the Einstein relation, which relates the mean FDT time and the fluctuations of the FDR time.

The problem of the mean FDR time was considered in \cite{Gruenbaum2013}. 
For quantum walks which are subject to repeated measurements, 
the return to the initial state and the transition to another state have quite different dynamical properties. 
First, both the return and the transition properties are very sensitive to the back-folded 
spectrum of Eq. (\ref{back_folded_spectrum}). The mean FDR time is topologically protected and equal to the number of 
non-degenerate back-folded eigenvalues in \cite{Gruenbaum2013,Yin2019}. We have not found such a topologically protected
time scale for the FDT properties. The mean FDT time is divergent near the degeneracies, in the presence 
of 1) a very small overlap $p_0=|\langle E_0|\psi_d\rangle|^2$, 2) merging of two phases, and 3) the big charge theory. We note that other scenarios for diverging mean FDT times can be found for example in the Zeno limit \cite{Dhar2015,Thiel2019quantization} and when three or more charges are merging  \cite{Yin2019}.
Another difference between the return and the transition problem is that, for instance, the total detection probability
$P_{\rm det}$ for the return is always unity, while the transition probability to another state is sensitive, e.g. to the geometric symmetry of the underlying graph \cite{Thiel2019b}. The qualitative difference between return and transition properties
originates in the fact that the return properties are based on the amplitude $u_n$ alone, whereas the transition 
properties depend on both amplitudes $u_n$ and $v_n$. This implies a more complex physical behavior for the transition 
properties. Although a unitary evolution without the projective measurements is already complex due to the different energy levels, 
the interruption by the measurement adds another timescale $\tau$ to the dynamics. This fact implies that the degeneracy of two or more of the phase factors affects the
dynamics substantially and that the dynamics depends strongly on the back-folded spectrum. It also explains why a
small coefficient $p_0$ has a similar effect: the effective dimensionality of the available Hilbert space is either reduced by the 
degeneracy of the phase factors or by the vanishing overlap $p_0$, since each phase factor carries the coefficient $p_0$ as
$p_0\exp(-i E_0\tau)$. This observation and the results of the calculations in Secs.\ref{sect:weak}, \ref{sect:merging} and \ref{sect:big}
can be summarized to the statement that divergent mean FDT times are caused by the proximity to a change of the effective Hilbert space 
dimensionality.

We have found that when a single stationary point, $z$ (or the pole $Z=1/z^*$) approaches the unit circle in the complex plane, we get a relation between the mean FDT time and the fluctuations of the FDR time, see Eqs. (\ref{n v weak},\ref{n v two charge}).  This is because the slow relaxations of $\phi_n$, which are controlled by a single pole $Z_i$, making all others irrelevant.

Our results also indicate that a quantum walk, interrupted by repeated measurements, is quite different from classical diffusion. The divergent mean FDT time reflects the fact that the transition to certain states can be strongly suppressed. In this sense the dynamics is controllable by choosing the time step $\tau$. This could be important for applications,
for instance, in a quantum search process: the search time for finding a certain quantum state depends significantly on the choice
of the time step $\tau$. Trapping of the quantum state by an external potential also influences strongly the value of the mean FDT time 
$\langle n \rangle$, as we have seen in our examples.

%How we localize the wave function with defects (or potential $U$) and sampling time $\tau$ can also give the Four %divergent mean FDT times that we discussed in the main text. Using Eq. (\ref{pk}), if one of the eigenstate %$|E_i\rangle\sim |\psi_{\rm d}\rangle$, which can be achieved by setting a defect on the latices node where we %put the detector, we have the big charge theory. This defect localizes the wave function at the detected state %$|\psi_{\rm d}\rangle$. Inversely, if we set the defect on the node of the initial state $|\psi_{\rm in}\rangle$ %(for instance on node $x=0$), from Eqs. (\ref{pk},\ref{qk}), we have $p_0\rightarrow 0$ and $|q_0|>>p_0$, this is %exactly the weak charge scenario. Putting the defect between the initial state and detected state will lead the %two merging charges case (or even more charges merge, but we can use the two merging charges case to approximate %the tendency). Finally, in the Zeno regime, the rapid detection attempts localize the wave function at the point %where it starts. We give a quantum speed limit for the state transform ($|\psi_{\rm in}\rangle\rightarrow %|\psi_{\rm d}\rangle$) in the Hilbert space. The beauty of our theory is that we start from a classical charge %picture, but in each circumstance we have clear quantum correspondence. 

\section{Acknowledgements}
We thank Felix Thiel and David Kessler, for many helpful discussions, which led to simplifications of some of the formulas of this paper.
The support of Israel Science Foundation's grant 1898/17 as well as the support by the Julian Schwinger Foundation (KZ) are acknowledged.

\appendix

\section{Order of ${\cal G}(z)$ and ${\cal D}(z)$ }\label{order}
\begin{figure*}
    \centering
    \includegraphics[width=1.6\columnwidth]{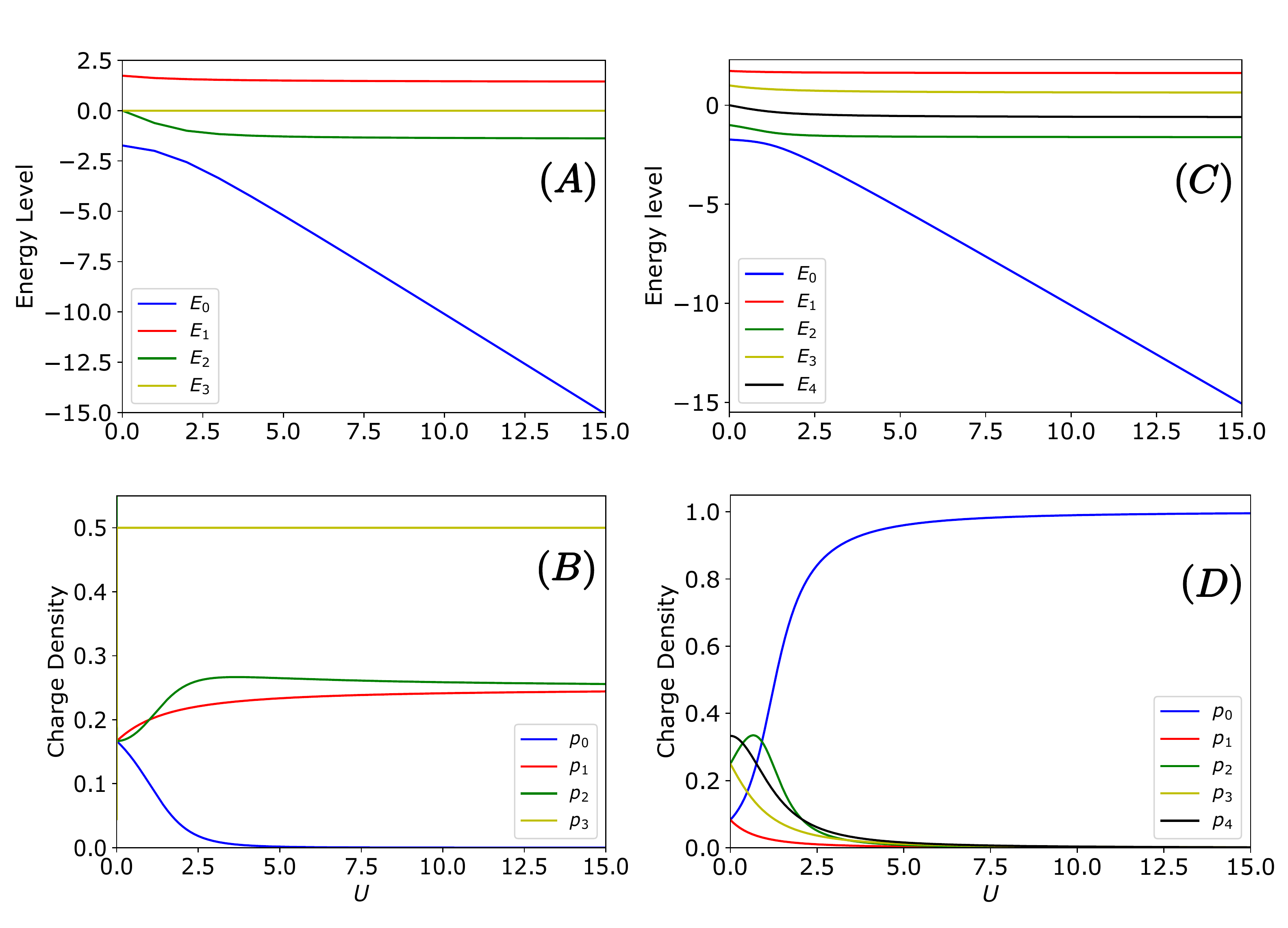}
    \caption{(A): the energy levels versus the potential $U$ of the Y-shaped molecule. (B): the energy levels versus the potential $U$ of the linear five-site molecule. (C): the charges $p_k$ versus the potential $U$ of the Y-shaped molecule. (D): the charges $p_k$ versus the potential $U$ of the linear five-site molecule.}
    \label{fig: energy level charge dnesity}
\end{figure*}
In this section we proof $\text{deg}({\cal D}(z))>\text{deg}({\cal G}(z))$ used in the main text.
Since $ {\cal G}(z) \simeq \sum q_i z^{w-1} + \cdots$, the highest order of ${\cal G}(z)$ is $(\sum_{i=0}^{w-1}q_i) z^{w-1}$. 
However, what is special in the transition problem is $\sum_{i=0}^{w-1}q_i=\langle \psi_{\rm d}|\psi_{\rm in}\rangle=0$, 
namely that the highest order vanishes, such that $\text{deg}({\cal G}) <w-1$ for the numerator.\\

Using Eq. (\ref{D}), ${\cal D}(z)\simeq \sum p_i e^{i E_i \tau} z^{w-1} + \cdots$ the leading order of $z$ is $\beta z^{w-1}=(\sum_{i=0}^{w-1}p_i e^{i E_i \tau}) z^{w-1}$.
\begin{equation}
    \sum_{i=0}^{w-1}p_i e^{i E_i \tau}=\langle \psi_{\rm d}|e^{i \hat{H}\tau}|\psi_{\rm d}\rangle\neq 0.
\end{equation}
Hence $\text{deg}({\cal D}(z))>\text{deg}({\cal G}(z))$.

\section{Weak charge}\label{weak charge Appendix}
In this section we derive Eqs. (\ref{Z_0},\ref{C_0},\ref{n weak}) of the main text. Following the same procedure, we can derive Eqs. (\ref{C_p},\ref{n two charges}) in the Sec. \ref{Two merging charges}.

As we mentioned in the main text, the weak charge $p_0\sim 0$ and the corresponding energy level is $E_0$. Using Eq. (\ref{Force field}), we have:
\begin{equation}
    0=\sum_{k=0}^{w-1}\frac{p_k}{e^{i\uptau E_k}-z}=\frac{p_0}{e^{i\uptau E_0}-z}+\sum_{k=1}^{w-1}\frac{p_k}{e^{i\uptau E_k}-z}.
    \label{for zeros}
\end{equation}
Assuming that $z_0=e^{i E_0\uptau}-\epsilon$. 
The $\epsilon$ is the first order approximation. Using Eq. (\ref{for zeros}), we have:
\begin{equation}
    0\approx \frac{p_0}{\epsilon}+\sum_{k=1}^{w-1}\frac{p_k}{e^{i\uptau E_k}-e^{i\uptau E_1}},
\end{equation}
hence
\begin{equation}
    \epsilon \sim  \frac{p_0}{\sum_{k=1}^{w-1}p_k/(e^{i\uptau E_1}-e^{i\uptau E_k})}.
    \label{epsilon}
\end{equation}
Using Eq. (\ref{relation of zero and pole}), the pole $Z_0$ in the main text reads:
\begin{equation}
    Z_0=\frac{1}{z_0^{\ast}}=\frac{1}{e^{-i\uptau E_0}-\epsilon^{\ast}}\sim e^{i\uptau E_0}(1+\epsilon^{\ast}e^{i\uptau E_0}) .
\end{equation}

The index $C_i$ is defined in Eq. (\ref{C_i general}). Plugging the pole $Z_0$ into Eq. (\ref{C_i general}), we obtain

\begin{align*}
    {\cal N}(Z_0) & \sim -e^{i E_0\tau}\bigg[\frac{q_0}{-\epsilon^{\ast}e^{2i E_0\tau}}+\sum_{j=1}^{w-1}\frac{q_j}{(e^{i E_j\tau}-e^{i E_0\tau})}\bigg]\\
    &\sim \frac{q_0}{\epsilon^{\ast}e^{i E_0\tau}},
\end{align*}
and
\begin{align*}
    {\cal D}^{\prime}(Z_0) & \sim \frac{p_0 e^{i E_0\tau}}{\epsilon^{\ast2}e^{4i E_0\tau}}+\sum_{j=1}^{w-1}\frac{p_j e^{i E_j\tau}}{(e^{i E_j\tau}-e^{i E_0\tau})^2}\\
    & \sim \frac{p_0 }{\epsilon^{\ast2}e^{3i E_0\tau}}.
\end{align*}
Hence the coefficient $C_0$ used in the main text reads:
\begin{equation}
    C_0\sim \frac{q_0}{p_0}\epsilon^{\ast}e^{2 i E_0 \tau}.
\end{equation}
Substituting $C_0$ and $Z_0$ into Eq. (\ref{weak 1}), the mean FDT time becomes
\begin{equation}
    \langle n \rangle \sim \frac{|C_0|^2}{(|Z_0|^2-1)^2} \sim \frac{|q_0|^2}{p_0^2}\frac{|\epsilon|^2}{(2Re[\epsilon* e^{-i E_0\tau}])^2}.
    \label{B6}
\end{equation}
Using the mathematical property $1/(1-\exp{[ix]})=1/2+i\cot{[x/2]}/2$ and the normalization condition $\sum_{k=1}^{w-1}p_k=1-p_0\sim 1$, we can simplify the parameter $\epsilon$. From Eq. (\ref{epsilon}), we have:
\begin{align*}
    \epsilon &\sim \frac{p_0}{\sum_{k=1}^{w-1}p_k/(e^{i\uptau E_0}-e^{i\uptau E_k})}\\
    & = e^{i E_0\tau}\frac{p_0}{\sum_{k=1}^{w-1}p_k/(1-e^{i\uptau (E_k-E_0)})}\\
    & = e^{i E_0\tau}\frac{2p_0}{\sum_{k=1}^{w-1}p_k(1+i\cot{[\uptau (E_k-E_0)/2]})}\\
    & \sim e^{i E_0\tau}\frac{2p_0}{1+i \sum_{k=1}^{w-1} p_k\cot{[\uptau (E_k-E_0)/2]}}.
\end{align*}
Plugging $\epsilon$ into Eq. (\ref{B6}), the mean FDT time reads
\begin{equation}
   \langle n \rangle \sim  \frac{|q_0|^2}{4p_0^2}\Bigg\{ 1+\bigg[\sum_{k=1}^{w-1} p_k \cot{[(E_k-E_0)\tau/2]}\bigg]^2\Bigg\}.
\end{equation}

\section{Two-charge pole $Z_p$\label{two charge pole}}
In this section we derive Eq. (\ref{Z_p}) of the main text.
When a pair of charges is nearly merging, say $\exp{(i E_a \tau)} \simeq \exp{(i E_b \tau)}$, one of the zeros denoted $z_p$ will be close to the unit circle. We define
$2\delta=(\Bar{E}_b-\Bar{E}_a)\uptau$, hence $\delta$ is a order parameter measuring this process. We first consider the two merging charges. Using Eq. (\ref{Force field}) we have
\begin{equation}
    \frac{p_a}{e^{i E_a\uptau}-z}=-\frac{p_b}{e^{i E_b\uptau}-z},
    \label{only two}
\end{equation}
which yields
\begin{equation}
    z_p^{(0)}=\frac{p_a e^{i E_b\uptau}+p_b e^{i E_a\uptau}}{p_a+p_b}.
\end{equation}
Now we take the background charges into consideration. 
\begin{equation}
    z_p=z_p^{(0)}-z_p^{(1)}.
\end{equation}
Plugging $z_p$ into Eq. \ref{Force field}, we have
\begin{widetext}
\begin{equation}
    0=
    \sum_{k=0}^{w-1}\frac{p_k}{e^{i E_k\uptau}-z}\approx\frac{p_a}{e^{i E_a\uptau}-z_p^{(0)}
    +z_p^{(1)}}+\frac{p_b}{e^{i E_b\uptau}-z_p^{(0)}
    +z_p^{(1)}}+\sum_{k\neq a,b}^{w-1}\frac{p_k}{e^{i E_k\uptau}-z_p^{(0)}} .
    \label{appendix two charges zero}
\end{equation}
The third part on the right-hand side is the effect of the background charges. We define it as $B$.
\begin{equation}
    B=\sum_{k\neq a,b}^{w-1}\frac{p_k}{e^{i E_k\uptau}-z_p^{(0)}}
    \approx \sum_{k\neq a,b}^{w-1}\frac{p_k}{e^{i E_k\uptau}-e^{i\uptau\frac{\Bar{E}_A+\Bar{E}_B}{2}}}.
    \label{B}
\end{equation}
Using Eqs. (\ref{appendix two charges zero},\ref{B}), we obtain
\begin{equation}
    z_p^{(1)}\sim\frac{B p_a p_b (e^{i E_a\uptau}-e^{i E_b\uptau})^2}{(p_a+p_b)^3+B(p_a^2-p_b^2)(e^{i E_a\uptau}-e^{i E_b\uptau})}\sim
    \frac{B p_a p_b (e^{i E_a\uptau}-e^{i E_b\uptau})^2}{(p_a+p_b)^3} .
\end{equation}
Since $e^{i E_B\uptau}-e^{i E_A\uptau}\sim \delta$, $z_p^{(1)}\sim \delta^2$. The background charges give only a second order effect 
$O(\delta^2)$ to the zero $z_p$ as we expected. Using Eq. (\ref{relation of zero and pole}) we have
\begin{equation}
    Z_p=Z_p^{(0)}+Z_p^{(1)}=\frac{1}{z_p^{\ast}}\approx \frac{p_A+p_B}{p_A e^{-i E_B\uptau}+p_B e^{-i E_A\uptau}}+\frac{B^{\ast}p_A p_B(e^{-i E_A\uptau}- e^{-i E_B\uptau})^2}{(p_A+p_B)(p_A e^{-i E_B\uptau}+p_B e^{-i E_A\uptau})^2} .
    \label{appendix Z_p}
\end{equation}
\end{widetext}

\end{document}